\documentclass[letterpaper,11pt]{article}

\usepackage[active]{srcltx}
\usepackage{amsrefs}

\usepackage{scrextend}
\usepackage{amsmath,amssymb,epsfig,bbm}
\usepackage{MnSymbol}
\usepackage{slashed}
\usepackage{amsthm}

\usepackage{float}

\usepackage{color}
\usepackage{mathrsfs}

\usepackage{dsfont}

\definecolor{co}{cmyk}{0,0.7,0.3,0}
\definecolor{darkgreen}{cmyk}{1,0,1,.2}
\definecolor{m}{rgb}{1,0.1,1}

\newcommand{\be}{\begin{equation}}
\newcommand{\ba}{\begin{eqnarray}}
\newcommand{\ea}{\end{eqnarray}}
\newcommand{\nn}{\nonumber}

\def\d{\delta}

\def\OO{\Omega}
\def\P{\Pi}

\def\ca{{\cal A}}
\def\cb{{\cal B}}

\def\cd{{\cal D}}
\def\ce{{\cal E}}
\def\cf{{\cal F}}
\def\cg{{\cal G}}
\def\ch{{\cal H}}

\def\ck{{\cal K}}

\def\co{{\cal O}}

\makeatletter
\newcommand{\eqnum}{\refstepcounter{equation}\textup{\tagform@{\theequation}}}
\makeatother

\newcommand{\pa}{\partial}

\newtheorem{thm}{Theorem}[subsection]

\newtheorem{conj}[thm]{Conjecture}
\newtheorem{definition}[thm]{Definition}

\newtheorem{remark}{Remark}

\newcommand{\cF}{{\cal F}}

\newtheorem*{definition*}{Definition}

\newcommand{\R}{\mathbb{ R}}

\fontfamily{yfrak}

\begin{document}

\vskip 25mm

\begin{center}

{\large\bfseries  

Constructing spectral triples over
holonomy-diffeomorphisms
and the problem of reconciling general relativity with quantum field theory
}

\vskip 5ex

Johannes \textsc{Aastrup}$^{a}$\footnote{email: \texttt{aastrup@math.uni-hannover.de}} \&
Jesper M\o ller \textsc{Grimstrup}$^{b}$\footnote{email: \texttt{jesper.grimstrup@gmail.com}}\\ 
\vskip 2ex

$^{a}\,$\textit{Mathematisches Institut, Universit\"at Hannover, \\ Welfengarten 1, 
D-30167 Hannover, Germany.}
\\[1ex]
$^{b}\,$\textit{Independent researcher.}
\\[2ex]

{\footnotesize\it This work is financially supported by entrepreneur Kasper Gevaldig, Copenhagen, Denmark, and by Master of Science in Engineering Vladimir Zakharov, Granada, Spain.}

\end{center}

\vskip 2ex

\begin{abstract}

\vspace{0.5cm}

In this paper we  
construct a candidate for a spectral triple on a quotient space of gauge connections modulo gauge transformations and show that it is 
related to a Kasparov type bi-module over two canonical algebras: the $\mathbf{HD}$-algebra, which is a non-commutative $C^*$-algebra generated by parallel transports along flows of vector fields, and an exterior algebra on a space of gauge transformations. The latter algebra is related to the ghost sector in a BRST quantisation scheme. Previously we have shown that key elements of bosonic and fermionic quantum field theory on a curved background emerge from a spectral triple of this type. In this paper we show that a dynamical metric on the underlying manifold also emerges from the construction. We first rigorously construct a Dirac type operator on the a quotient space of gauge connections modulo gauge transformations, and discuss the commutator between this Dirac type operator and the $\mathbf{HD}$-algebra.
To do this we first construct a gauge-covariant metric on the configuration space and use it to construct the  triple. The key step in this construction is to require the volume of the quotient space to be finite, which amounts to an ultra-violet regularisation. Since the metric on the configuration space is dynamical with respect to the time-evolution generated by the Dirac type operator in the  triple, it is possible to interpret the regularisation as a physical feature (as opposed to static regularisations, which are always computational artefacts). Finally, we construct a Bott-Dirac operator that connects our construction with quantum Yang-Mills theory.

\end{abstract}

\newpage
\tableofcontents

\section{Introduction}

Arguably the two most important problems in theoretical high-energy physics are the question of how to reconcile general relativity with quantum theory and the question of how to formulate non-perturbative quantum field theory rigorously. Both of these problems are intrinsically related to the ultra-violet limit and thus it is reasonable to assume that they are closely related.
Indeed, this assumption is the starting point of the research project that we report on in this paper: that general relativity could emerge from a rigorous formulation of non-perturbative quantum field theory. 

Here is why this makes sense: a non-perturbative theory of quantum field theory will most likely require an ultra-violet regularisation, which must be gauge covariant. Gauge covariance implies, however, that the regularisation depends on the gauge field, i.e. it will not be a constant throughout the configuration space of gauge connections. This, in turn, means that it will be subjected to the time-evolution generated by the Hamilton operator in that theory \cite{Aastrup:2020jcf}. But a dynamical ultra-violet regularisation can, in contrast to a static regularisation, be interpreted as a physical feature and with that we arrive at the crux of this argument: such an ultra-violet regularisation will necessarily encode metric information about the underlying three-dimensional manifold and since the regularisation is dynamical so will the spatial metric on that manifold be. And hence the first ingredient of general relativity emerges from a non-perturbative theory of quantum field theory: a dynamical metric on a three-dimensional manifold.

To explain how we approach this idea let us take a step back and begin with the question of how to reconcile general relativity with quantum theory. This problem has been known for almost a century and the strategy has almost always been to apply the framework of the latter to the former. That is, to construct a theory of quantum gravity. There exist, however, a second logical possibility, which is to apply the framework of general relativity to quantum theory. It turns out that this idea leads to a construction that addresses not only the problem of non-perturbative quantum field theory but also involves a dynamical ultra-violet regularisation of the kind that we just discussed.

Concretely, the idea is to construct a dynamical geometry on a configuration space $\ca$ of gauge connections via an application of noncommutative geometry \cite{Aastrup:2005yk}. In noncommutative geometry the geometry of a space is encoded in a spectral triple, that consist of a $C^*$-algebra, a Dirac operator and a Hilbert space \cite{Connes:2008vs,Connes:1996gi,ConnesBook}. To construct a spectral triple over the configuration space $\ca$ we begin with an algebra that encodes how tensor degrees of freedom are translated on a three-dimensional manifold $M$. This is the $\mathbf{HD}$-algebra \cite{AGnew}, which is generated by parallel transports along flows of vector fields. The $\mathbf{HD}$-algebra can be interpreted as a noncommutative algebra of functions on $\ca$.
Once we have the $\mathbf{HD}$-algebra we complete the triple by constructing a Dirac operator and a Hilbert space. Since the configuration space is infinite-dimensional the Dirac operator will also be infinite-dimensional just as the Hilbert space will involve an integration over the configuration space, i.e. a type of path integral.

In previous publications \cite{Aastrup:2017atr,Aastrup:2019yui,Aastrup:2020jcf} we have shown that a spectral triple of this type encodes the basic ingredients of a Yang-Mills quantum field theory coupled to a fermionic field on a curved background. The Yang-Mills and the Dirac Hamiltonians emerge from the square of a Bott-Dirac operator, which is a natural extension of the Dirac operator, and the canonical commutation and anti-commutation relations are encoded in the interaction between the $\mathbf{HD}$-algebra and the Bott-Dirac operator. In this paper we show how a dynamical metric on the underlying manifold also emerges from a spectral triple construction of this type in a flat and local limit (flat and local with respect to the quotient space).


 This means that the framework, which we present, involves several layers of unification. First there is a unification between bosonic and fermionic quantum field operators, where the fermions emerge from the CAR algebra that is used to construct the Dirac operator, secondly there is an emergence of additional degrees of freedom stemming from inner automorphisms of the $\mathbf{HD}$-algebra. In the case of the standard model it is inner automorphisms of an almost commutative algebra that give rise to the gauge and Higgs sectors \cite{Connes:2006qj,Chamseddine:1991qh,Chamseddine:2006ep}. Thirdly, there is the emergence of a dynamical metric field on the underlying manifold, which means that general relativity could emerge from this framework too.   
Note that these three elements of unification all take place at a 'pre-QFT level', which is deeper than other types of unification such as supersymmetry, grand unification and the unification found in string theory.

A key purpose of the present paper is to make  the construction of the  spectral triple rigorous. This means that we show that a Hilbert space exists in which the Dirac operator and the $\mathbf{HD}$-algebra exist as operators. In order to achieve this we first construct a gauge covariant metric on $\ca$. We construct the metric so that the volume of the quotient space of gauge fields modulo gauge transformations is finite. 

In fact, it is the finite-volume  that leads us to the dynamical, ultra-violet regularisation previously mentioned.
The point is that for the quotient space to have a finite volume it is necessary that the metric regularises the integration over the configuration space needed to construct the Hilbert space. Such an ultra-violet regularisation would normally be interpreted as a computational artefact, but, as already said, the additional requirement of gauge-covariance implies that it is dynamical, which in turn permits us to interpret it as a physical feature.
 To the best of our knowledge a dynamical ultra-violet regularisation has never been studied before.

Concretely, we use the gauge covariant metric on $\ca$ to construct first a Hilbert space on the quotient space of gauge connections moduli gauge transformations. We then show that a candidate for a spectral triple exist on this quotient space. Next we outline how a BRST procedure \cite{Becchi:1975nq,Tyutin:1975qk,Barnich:2000zw,Henneaux:1992ig} can be used to extend this construction to all of $\ca$. Finally we find that our framework emerges from an unbounded Kasparov type bi-module over the $\mathbf{HD}$-algebra and an exterior algebra of functions on the space of gauge transformations. Here the BRST operator works like an exterior derivative with respect to the exterior algebra. Note that the two algebras in this bi-module are highly canonical as they only depend on a choice of gauge group, the dimension of space, as well as a metric dependency in the ghost sector, which we expect to be unphysical. This shows that our framework comes with a high level of canonicity.

Concerning canonicity it is important to state that the strategy in this paper is to construct a concrete realisation of the spectral triple construction, i.e. to show that there exist {\it a} metric on $\ca$ for which the spectral triple construction exists rigorously. The metric that we construct is somewhat contrived, but this is unimportant since the purpose is simply to demonstrate that our construction can be realised concretely.

The spectral triple that we find comes with a number of important caveats: 
first, the construction of the metric relies on a gauge fixing procedure. This is, as already said, in itself not a problem, but it means that for the triple to exist in the non-Abelian case it requires the resolution of the Gribov ambiguity \cite{Gribov:1977wm,Singer:1978dk}, which we have not solved. Secondly, regardless of the specificity of the metric the entire construction of the spectral triple does depend on a resolution of the Gribov ambiguity since it is build on the quotient space $\cf$.  Thirdly, we are not yet able to prove that the triple we have found satisfies all the requirements of a spectral triple. In particular, we construct two different metrics on the configuration space $\cf$. In the case of the first metric the Dirac operator is self adjoint but the commutator with the $\mathbf{HD}$-algebra is not bounded. In the second case the commutator is bounded but we are unable to prove that the Dirac operator is self adjoint. Also, we have not proven that the Dirac operator has compact resolvent. Nevertheless, we permit ourselves a slight misuse of terminology and use the term spectral triple.

The philosophy behind this research project, which was first commenced in \cite{Aastrup:2005yk}, is to found a fundamental theory on the mathematics of empty space. A candidate for a fundamental theory must be highly irreducible in terms of further scientific reductions and the way to secure that is to base it on something that is conceptually almost empty. The construction that we propose depends primarily  on the $\mathbf{HD}$-algebra, which essentially encodes how 'stuff' is moved around in empty space, combined with a metric principle, and the choice of gauge group and dimension of space. Concerning the gauge group there are two natural choices: either $SO(3)$, which corresponds to Levi-Civita connections, or $SU(2)$, which corresponds to spin-connections. But the construction makes sense for a general gauge group $G$ that corresponds to gauge fields.

This construction can be seen a natural continuation of Einstein's two theories of relativity: what we propose is a theory of relativity on a configuration space related to a three-dimensional manifold. In order to emphasise this we propose the name\footnote{We have previously used the term "quantum holonomy theory". We are indebted to Jarl Sidelmann for suggesting a better name.} 'configurational relativity'.

Finally, let us mention the notion of a distance on a configuration space of gauge connections is not new but
was discussed already by Feynman \cite{Feynman:1981ss} and Singer \cite{Singer:1981xw} (see also \cite{Orland:1996hm} and references therein). What is new in our approach is that we do not consider only the $L^2$-metric on the configuration space but a large class of covariant metrics. This is what opens the door to a unifying picture that ties fermionic and bosonic quantum field theory together in a novel way.\\

The paper is organized as follows: We begin with a construction of a metric on the configuration space $\ca$ in section 2. In section 3 we then construct a Hilbert space in which we represent the $\mathbf{HD}$-algebra, which we introduce in section 4. We construct a Dirac operator in section 5 and analyse its commutators with the $\mathbf{HD}$-algebra. In section 6 we then discuss an alternative construction of a metric on $\ca$. Next, we analyse the ultra-violet limit of Hilbert space states in section 7. Until this point we have worked on a gauge fixed 'slice' of $\ca$, but in section 8 we outline the construction of a BRST operator alongside a ghost-sector, which permits us to work on the entire $\ca$. This step automatically leads us to discuss Kasparov bi-modules. In section 9 we then introduce the Bott-Dirac operator, which puts us in contact to perturbative quantum field theory. Finally, in section 10, we explain why a dynamical metric on $M$ emerges in a local limit. We end with a discussion in section 11.

\subsection{Outline of the construction}

We begin with a broad outline of 
our construction. Thus, the following is merely a sketch meant to provide the reader with an overview of our approach.\\

The starting point is the construction of a gauge covariant metric $g_\cf$ on a configuration space $\ca$ of gauge fields. The central requirement is that with this metric the volume of $\cf$, which is the quotient space of gauge connections modulo gauge transformations, is finite
$$
\mbox{Vol}_{g_\cf}(\cf) < \infty.
$$
This permits us to construct a Hilbert space $\ch$ with an inner product
$$
\langle \eta, \zeta\rangle = \int_\cf [d\nabla] \mbox{det}(g_\cf)   \overline{\eta}(\nabla)\zeta(\nabla), 
$$
where $\eta$ and $\zeta$ are functions on $\cf$, where $[d\nabla]$ is a measure on $\ca$ and $\mbox{det}(g_\cf)$ the determinant of $g_\cf$. 
If we let $\{\varphi_i\}$ be an orthonormal basis of one-forms in $\OO^1(M,\mathfrak{g})$, where $\mathfrak{g}$ is the Lie algebra of the gauge group $G$, then we can write $\nabla\in\cf$ as
\begin{equation}
\nabla = \nabla_0 + \sum_i x_i \varphi_i
\label{xxxx}
\end{equation}
where $\nabla_0$ is an arbitrary basepoint in $\cf$. With this we can write the measure $[d\nabla]$ as\footnote{We will only consider the $x$-directions along $\cf$.}
$$
[d\nabla] = dx_1 dx_2 \ldots dx_n \ldots 
$$
while the determinant of $g$ 
serves as a dampening factor that secures convergence. 
Next we consider translations on $\ca$. Since two arbitrary connections $\nabla$ and $\nabla'$ always differ by a one-form $\omega$
$$
\nabla' = \nabla+\omega,
$$
this corresponds to a translation operator $U_\omega$
$$
U_\omega \xi(\nabla) = \xi(\nabla+\omega)
$$
on functions $\xi$ on $\ca$. 
If we consider infinitesimal translations 
$
 \frac{\pa}{\pa x_i}
$ 
(see (\ref{xxxx})) then we can construct a Bott-Dirac type operator on $\cf$, which has the form \cite{Aastrup:2019yui}
$$
B= \sum_{i=1}^\infty \left( \bar{c}_i \frac{\pa}{\pa x_i} + {c}_i F_i \right),
$$
where $F_i$ is the curvature of $\varphi_i$ and where $(c_i,\bar{c}_i)$ are elements in an infinite-dimensional Clifford algebra. The square of $B$ gives us then the Hamiltonian of a Yang-Mills theory coupled to a fermionic field
$$
B^2 = \sum_{i=1}^\infty \left( \left(\frac{\pa}{\pa x_i}\right)^2 +  \left( F_i\right)^2 \right) + "\mbox{fermionic terms}"
$$
in a form, which resembles an infinite-dimensional harmonic oscillator. Here the fermionic sector emerges from the infinite-dimensional Clifford algebra, which is required in order to construct $B$, i.e.
$$
\{ c_i , {c}_j \} = \d_{ij},\quad \{ \bar{c}_i , \bar{c}_j \} = -\d_{ij},\quad \{ c_i , \bar{c}_j \} = 0,
$$
gives rise to the canonical anti-commutation relations of a fermionic field operator \cite{Aastrup:2020jcf}
$$
\{  {\psi}^\dagger({\bf m}_1) , {\psi}({\bf m}_2) \} = \d({\bf m}_1-{\bf m}_2).
$$
Furthermore, the state
$$
\Psi(A) = e^{i CS(A)}
$$
where
$$
CS(A) = \int_M \mbox{Tr} \left( A\wedge dA + \frac{2}{3} A\wedge A \wedge A \right)
$$
is the Chern-Simons functional, will lie in the kernel of $B$, i.e.
$$
B \Psi(A) =0,
$$
which thus gives us the ground state of the theory \cite{Aastrup:2020jcf}.

Note that with this Bott-Dirac operator the fermions will a priori be one-forms, i.e. there will be a one-to-one correspondence between the bosonic and the fermionic sectors, which is at odds with special relativity and the spin-statistics theorem. 
In \cite{Aastrup:2020jcf} we found one possible solution this problem, which involved a modified Bott-Dirac operator.

To summarise, we find that the Bott-Dirac operator, which interacts with the ${\mathbf{ HD}}$-algebra generated by parallel transports, forms a type of noncommutative spectral triple
$$
(B, {\mathbf{ HD}}, \ch )
$$
over $\cf$,
which encodes the basic ingredients of a Yang-Mills-Dirac quantum field theory on a curved background.

\section{A metric on a configuration space of gauge connections}

 In this section we construct a gauge-covariant metric on the configuration space $\ca$. In a first step we construct the metric on a gauge-fixed slice $\cf$ of $\ca$ for then to extend it in a covariant manner to all of $\ca$. 
 The key property of the metric is that it renders $\cf$ with a finite volume. \\
 
Let $\ca$ be the space of smooth $G$-connections, where $G$ is the gauge group that acts on $\ca$. A gauge fixing of $\ca$ is a subset $\cf$ of $\ca$ with the property that for each $\nabla \in \ca$ there is a $g\in G$ with $g(\nabla )\in \cf$. We will for the  purpose of this paper assume that for each $\nabla \in \ca$ there is exactly  one $g\in G$  with $g(\nabla )\in \cf$. 
\begin{remark}
This assumption presupposes the absence of the Gribov-ambiguity \cite{Gribov:1977wm,Singer:1978dk}. In this paper we ignore the Gribov ambiguity, which means that our results hold in the Abelian case and in the case where the Gribov ambiguity has been resolved. We shall comment on the Gribov ambiguity in the final section. 
\end{remark}
\begin{remark}
Note that in the absence of the Gribov ambiguity the gauge fixing $\cf$ is equal to the quotient space $\ca/\cg$ where $\cg$ is the space of gauge transformations. Throughout the paper we shall occasionally refer to $\cf$ as the quotient space. It is understood that this terminology is correct only in the absence of the Gribov ambiguity.
\end{remark}
To construct a gauge covariant entity on $\ca$ it suffices to construct it on $\cf$, and then use the action of $\cg$ to extend it to all of $\ca$.
We will demonstrate this now  by constructing a gauge covariant metric on $\ca$.

First, by choosing a $\nabla_0 \in \ca$ we can identify $\ca$ with $\Omega^1(M,\mathfrak{g})$, where $\mathfrak{g}$ denotes the Lie-algebra of $\cg$, by writing a given connection $\nabla \in \ca$ as
$$\nabla=\nabla_0+A, \quad A\in \Omega^1(M,\mathfrak{g}). $$
Next we choose the Hodge-Laplace operator $\Delta=dd^*+d^*d$ acting on $\Omega^1(M,\mathfrak{g})$. Hodge theory gives an $L^2$-decomposition 
$$\Omega^1(M,\mathfrak{g})=d(\Omega^0(M,\mathfrak{g}))\oplus  d^*(\Omega^2(M,\mathfrak{g}))\oplus H^1(M,\mathfrak{g}) $$
where $d^* :\Omega^*(M,\mathfrak{g}) \to \Omega^{*-1}(M,\mathfrak{g})$ is the adjoint of $d$ and   
$$H^1(M,\mathfrak{g}) =\hbox{Ker}\{  \Delta :\Omega^1(M,\mathfrak{g})\to \Omega^1(M,\mathfrak{g})  \}.$$
The gauge fixing, which we want to consider for the construction of the metric, is 
$$\cf=d^* (\Omega^2 (M,\mathfrak{g})) .$$
Modulo $H^1(M,\mathfrak{g})$ this corresponds to the gauge fixing $d^*A=0$, which in case of the flat metric on $M$ corresponds to $\partial^\mu A_\mu=0$.

Let $\{ \phi_i\}$ be a basis of eigenvectors of $\Delta$, i.e. $\Delta \phi_i=\lambda \phi_i$. Note that if $\lambda_i\not= 0$, we can split $\phi_i=\phi_{i_1}+\phi_{i_2}$ with $\phi_{i_1}\in d(\Omega^0(M,\mathfrak{g}))$, $\varphi_{i_2}\in d^*(\Omega^2(M,\mathfrak{g}))$ and such that both $\varphi_{i_1}$ and $\varphi_{i_2}$ fulfill
\begin{equation*}
    \Delta \phi_{i_1}=\lambda_i \phi_{i_1} \hbox{ and }\Delta \phi_{i_2}=\lambda_i \phi_{i_2} . 
\end{equation*}
Consequently we can choose a basis $\{  \varphi_i \}$ for $\cf$ consisting of eigenvectors of $\Delta$.

We now construct a preliminary metric on $T\cf$, which we need in order to construct the final metric. We can choose the basis $\{ \varphi_i\}$ such that this is orthonormal with respect to the norm 
\begin{equation}
\langle \xi , \eta \rangle_p=\langle (1+\tau\Delta)^p \xi, (1+\tau\Delta)^p \eta \rangle,  
\label{regn}
\end{equation}
where $p$ and is a positive real number and where $\tau$ is a coefficient. Furthermore we choose coordinates $(x_1,x_2,x_3,\ldots )$ according to $x_1\varphi_1+x_2\varphi_2+\ldots $. Finally, in a point 
$$x=(x_1,x_2,x_3,\ldots) \in \cF $$
we can identify $T_x\cF$ with $\cF$ itself, which means that we here also have the coordinates 
$$T_x\cf\ni \xi=(\xi_1,\xi_2,\xi_3,\ldots )=\xi_1\varphi_1+\xi_2\varphi_2+\xi_3\varphi_3+\ldots $$

In particular in these coordinates we have
$$\langle \xi , \eta    \rangle_p =\sum_{i=1}^\infty \xi_i\bar{\eta}_i .$$

\begin{remark}
The Sobolev-type norm (\ref{regn}) plays the role of an ultra-violet regularisation, which is needed in order to obtain a representation of the $\mathbf{HD}(M)$-algebra, which we will introduce shortly.
\end{remark}

\subsection{A finite dimensional metric}

We will now construct a Riemannian metric $\langle \cdot , \cdot \rangle_{\mathbb{S}^{n}} $ on $\mathbb{R}^n$ with the following properties:
\begin{enumerate}
    \item The metric is invariant under the action of $O(n-1)$.
    \item The determinant of the metric is $\pi^{-n}e^{-(x_1^2+x_2^2+\ldots +x_n^2)}$.
    \item On $\mathbb{S}^{n+1}$, the one-point compactification of $\mathbb{R}^n$, the metric extends to a smooth metric.  
    \item Asymptotically we have 
    $$ \langle \partial_i,\partial_i  \rangle_{\mathbb{S}^{n}} (x) = O( (x_1^2+x_2^2+\ldots +x_n^2)^ke^{-\frac{1}{n-1}(x_1^2+x_2^2+\ldots +x_n^2)}),  $$
    with $x=(x_1,x_2,\ldots ,x_n)\in \mathbb{R}^n$, for some power of $k$.
\end{enumerate}

We construct this metric by mapping $\R^n$ to $\mathbb{S}^n$ and pull back the metric from $\mathbb{S}^n$ to $\R^n$. Under this map we want $\mathbb{S}^n$ to be the one point compactification of $\R^n$. In particular since
$$ \int_{\R^n} \pi^{-\frac{n}{2}}e^{-(x_1^2+x_2^2+\ldots +x_n^2)} dx_1\cdots dx_n =1$$
we equip $\mathbb{S}^n$ with the standard metric scaled so that $\mathbb{S}^n$  has volume $1$.

We first apply polar coordinates 
$$
\mathbb{R}^n\rightarrow [ 0,\infty[ \;\times\; \mathbb{S}^{n-1}
$$
In polar coordinates the determinant of the metric should be $\pi^{-\frac{n}{2}} r^{n-1} e^{-r^2}$.
If we consider the sphere $\mathbb{S}^n$ as the unit sphere in $\R^{n+1}$ we can identify it as 
$$ \mathbb{S}^{n} \simeq [-1,1] \;\times\; \mathbb{S}^{n-1}  $$
where the set $\{-1\}\times \mathbb{S}^{n-1} $ is degenerate to a point, and the same with  $\{1\}\times \mathbb{S}^{n-1} $.

We want to construct our metric with a mapping  
$$
[ 0,\infty[\; \times\; \mathbb{S}^{n-1} \;\stackrel{\phi}{\longrightarrow}\; \mathbb{S}^{n} \simeq [-1,1] \;\times\; \mathbb{S}^{n-1}
$$
of the form: $\phi(r,s)= (y(r),s)$, $y(0)=-1$. 

Let $g_{sp}$ be the standard metric on $\mathbb{S}^n$ so that the volume is $1$. The requirement on $\phi$ is, that if we take the pull back of $g_{sp}$ with respect to $\phi$ the determinant of the metric should be  $\pi^{-\frac{n}{2}}e^{-(x_1^2+x_2^2+\ldots +x_n^2)}$, i.e. 
$$\hbox{det} (\phi^*(g_{sp}))=\pi^{-\frac{n}{2}}e^{-(x_1^2+x_2^2+\ldots +x_n^2)}. $$
This will give us an ordinary differential equation for $y(r)$ with the initial value condition $y(0)=-1$. The solution to this initial value problem  will, due to the requirement on the volume of $\mathbb{S}^n$, give a bijective map 
$$\phi : [ 0,\infty[\; \times\; \mathbb{S}^{n-1} \;\longrightarrow\;  [-1,1) \;\times\; \mathbb{S}^{n-1} ,$$
and the pulled backed metric will by definition have the desired determinant. Also the invariance under $O(n)$ and that $\mathbb{S}^n$ is the one-point compactification of $\R^n$ is by construction. 
\\

To see the asymptotics of the metric let us write the differential equation concretely:

Let $g_{s}$ denote the standard metric on $\mathbb{S}^{n-1}$. 
The standard metric on $\mathbb{S}^{n}$ is:
$$
\frac{1}{1-y^2} \vert \pa_y \rangle \langle \pa_y \vert + (1-y^2) g_{s}
$$
since on a point $y\in [-1,1]$ the radius of the sphere is $\sqrt{1-y^2}$, and if we take a curve $y\rightarrow (y,\sqrt{1-y^2})$ then the derivative is $(1,\frac{-y}{\sqrt{1-y^2}})$ with length
$$
1+ \frac{y^2}{1-y^2} = \frac{1}{1-y^2}.
$$
Since we want $\mathbb{S}^{n}$ to have volume $1$ we consider the metric
$$
T\left(\frac{1}{1-y^2} \vert \pa_y \rangle \langle \pa_y \vert + (1-y^2) g_{s}\right)
$$
where $T=(\hbox{vol}(\mathbb{S}^{n}))^{-\frac{2}{n}}$.
We will pull back this metric via $\phi$:
$$
\phi'(\pa_r) = y' \pa_y\qquad \phi'(\pa_\theta) = \pa_\theta.
$$
Thus the pulled back metric is given by
$$
T\left( \frac{(y')^2}{1-y^2} \vert \pa_r \rangle \langle \pa_r \vert + (1-y^2) g_{s} \right).
$$
The determinant of this metric is
$$
T^n\frac{(y')^2}{1-y^2} (1-y^2)^{n-2} =T^n (y')^2(1-y^2)^{n-3}
$$
and we thus get the equation
\begin{equation}
T^n(y')^2(1-y^2)^{n-3} =\pi^{-\frac{n}{2}} r^{n-1}e^{-r^2}
\label{zzzz}
\end{equation}
leading to the differential equation
$$
y' = C \frac{r^{\frac{n-1}{2}} e^{-\frac{r^2}{2}}}{(1-y^2)^{\frac{n-3}{2}}},
$$
where $C$ is some constant. The solution to this differential equation is given by the equation
$$
\int_{-1}^y (1-x^2)^{\frac{n-3}{2}} dx =  C\int_0^r t^{\frac{n-3}{2}} e^{-\frac{t^2}{2}} dt.
$$
For $y \to 1$ the left hand side has the expansion $D-(1-y)^{\frac{n-1}{2}}$ for some constant $D$, and the right hand side has the asymptotic expansion 
for $r\to \infty$ given by $D-r^{\frac{n-3}{2}} e^{-\frac{r^2}{2}}$. Asymptotically we thus get 
$$1-y \sim  r^{\frac{n-3}{n-1}} e^{-\frac{r^2}{n-1}}.$$
Since $\phi' (\partial_r)=y'\partial y$ we thus get 
$$\langle \partial_r ,\partial_r \rangle \sim  \frac{(y')^2}{1-y^2} \sim \frac{r^{\frac{8}{n-1}}e^{-\frac{2r^2}{n-1}}}{r^{\frac{n-3}{n-1}} e^{-\frac{r^2}{n-1}}} =r^{\frac{8}{n-1}}e^{-\frac{r^2}{n-1}} .$$
From here the asymptotics of $\langle \partial_r ,\partial_r \rangle$ follows. From the invariance under $O(n-1)$ and the determinant being $\pi^{-n}e^{-(x_1^2+x_2^2+\ldots +x_n^2)}$ follows the asymptotics of the rest of the coordinates.

\subsection{The metric on $T\cF$}

We are now ready to construct the final metric on $T\cF$. For a point $$x=(x_1,x_2,x_3,\ldots) \in \cf$$ 
we have the coordinates $(\xi_1,\xi_2,\xi_3,\ldots)$ on $T_x\cF$. Let $\{ V_j\}$ be the eigenspaces for $\Delta $ intersected with $\cF$, i.e. if $E_j$ is the eigenspace of $\lambda_j$, we have $V_j=E_j \cap \cF$. We assume that the eigenspaces are arranged in ascending order with respect to the eigenvalues of the eigenspaces. Let $m\in \mathbb{N}$ be a fixed chosen number. $m$ will be the minimal dimension of $\mathbb{S}^{n+1}$, which we will be using in the construction.   We will bundle the eigenspaces $\{ V_j\}$ together in subspaces $W_l$ such that each subspace $V_i$ is contained in a $W_l$, and such that the dimension of each $W_l$ is bigger than $m$. If for example $\hbox{dim} (V_1)=1$, $\hbox{dim} (V_2)=5$ and $m=3$, then $W_1=V_1\oplus V_2 $. If $\hbox{dim} (V_1)=5$, then $W_1=V_1 $. 

We will relabel the basis for $\cF$ as $\{ \varphi_{l_j}\}$, where $\varphi_{l_j}\in W_l$ for all $j$. We also write the coordinates as $(x_{l_j})$ and $(\xi_{l_j})$.
For given vectors $(\xi_{l_j}),(\eta_{l_j})\in T_{(x_{l_j})} \cF$ we define
$$  \langle (\xi_{l_j})    , (\eta_{l_j})  \rangle_m = \sum_{l} \langle (\xi_{l_1}, \ldots ,\xi_{l_{dim ( W_l)}}) , (\eta_{l_1}, \ldots ,\eta_{l_{dim ( W_l)}}) \rangle_{\mathbb{S}^{dim (W_l)+1}} .$$
Note that due to the coupling of the eigenspaces of $\Delta$ and the invariance of the metrics $\langle \cdot ,\cdot \rangle_{S^{k+1}}$, the metric $\langle \cdot ,\cdot \rangle_m$ is an invariant of $\Delta$, and therefore an invariant of the metric chosen on the underlying manifold.

Let $T_\cf\ca$ denote the bundle $T\ca$ restricted to $\cf$. We can extend the metric $\langle \cdot ,\cdot \rangle_m$ to  $T_\cf\ca$ be setting it equal to $\langle \cdot , \cdot \rangle_p$ on the directions orthogonal to $T\cf$. We will also denote this metric by $\langle \cdot ,\cdot \rangle_m$.

Once we have a metric on $T_\cf\ca$ we can extend to entire $T\ca$ in the following way: For a $\nabla \in \ca$ there is a unique\footnote{ Here we once more assume the absence of the Gribov ambiguity.} $g\in \cg$ with $g(\nabla )\in \cf$. For two vectors in $\xi,\eta T_\nabla \ca$ we define
\begin{equation}
\langle \xi , \eta  \rangle_m = \langle g_*(\xi) , g_*(\eta)  \rangle_m  ,
\label{Georgia}
\end{equation}
where $g_*:T_\nabla \ca \to T_{g(\nabla)}\ca$ is the differential of $g$.

\begin{remark}
The metric $\langle \cdot ,\cdot \rangle_m$ has a few key characteristics. First of all, if we compute its determinant it is on $\cf$ by construction given by 
$$\prod_i^\infty \frac{1}{\sqrt{\pi}}e^{-x_i^2} ,$$
where the product is taken over all directions in $\cf$. It hence has volume $1$. The metric is however not complete since on each $W_l$ it compactifies to a sphere.  

Secondly, note how $\langle \cdot ,\cdot \rangle_m$ is constructed with respect to a basis, that is orthonormal with respect to the Sobolev norm (\ref{regn}). This is significant since it means that the ultra-violet regularisation, which the Sobolev norm represents, will not be encoded in the determinant of the metric or in the spectrum of the Dirac operator, which we shall construct in section \ref{Diracoperator}. If we had built the ultra-violet regularisation into the metric $\langle \cdot ,\cdot \rangle_m$ in a more direct fashion the volume of $\cf$ would be zero. We shall discuss this issue in the final discussion.

Thirdly, the metric $\langle \cdot ,\cdot \rangle_m$ depends on a choice of gauge fixing. This is, however, not a problem because the purpose in this paper is simply to provide one example of a metric, for which the ensuing spectral triple construction can be realised concretely. 
\end{remark}

\section{The Hilbert space}

In this section we construct a Hilbert space $L^2(\cf)$. The central feature of this Hilbert space is that it is regularised by a dampening factor given by the determinant of the metric $\langle \cdot    , \cdot  \rangle_m$ and by the use of $\langle \cdot , \cdot \rangle_p$ in (\ref{regn}).\\

Thus, the task is to construct the Hilbert space $L^2(\cF)$ with respect to the metric $\langle \cdot ,\cdot \rangle_m$. We construct this as an inductive limit of Hilbert spaces: For each $k$ consider 
$$\ch_k=L^2(\R^k,\pi^{-\frac{k}{2}}e^{-(x_1^2+x_2^2+\ldots +x_k^2)}dx_1\cdots dx_k).$$
We can identify 
$$\cF_k:=\{x_1\varphi_1+\ldots + x_k\varphi_k| x_1,\ldots ,x_k\in \R\},$$
with $\R^k$ simply by mapping 
$$(x_1,\ldots , x_k) \to  x_1\varphi_1+\ldots + x_k\varphi_k ,$$
and we can define $L^2(\cF_k)$ as $\ch_k$. 

The mapping $ \iota_{k,k+1}:\ch_k\to \ch_{k+1}$ given by
$$ (\iota_{k,k+1})(\xi)(x_1,\ldots , x_k,x_{k+1})=\xi(x_1,\ldots , x_k)$$
is an embedding of Hilbert spaces. We can hence define
\begin{definition}
The Hilbert space $L^2(\cF)$ is defined as the inductive limit of the system of Hilbert spaces $(\ch_k,\iota_{k,k+1})_{k\in \mathbb{N}}$. 
\end{definition}

Next we want to construct the Hilbert space $L^2(\cF,\bigwedge^* T\cF )$. For this we again label the basis $\{\varphi_{l_j} \}$ and $\{ \xi_{l_j} \}$. We can identify each $W_l$ with $\R^{dim(W_l)}$ via
$$(x_1,\ldots ,x_{dim(W_l)})\to x_1\varphi_1+\ldots +x_{dim(W_l)} \varphi_{dim(W_l)},$$
and then consider $W_l$  with the metric $\langle \cdot ,\cdot \rangle_{\mathbb{S}^{dim(W_l)+1}}$. This gives us 
$$L^2(W_l,\bigwedge^* TW_l ).$$ Note that 
$$L^2(W_l\oplus W_k,\bigwedge^* T (W_l\oplus W_k) ) =L^2(W_l,\bigwedge^* TW_l ) \hat{\otimes} L^2(W_k,\bigwedge^* TW_k ),$$
and the map 
$$\iota_{l,l+k}:L^2(W_l,\bigwedge^* TW_l ) \to L^2(W_l\oplus W_k,\bigwedge^* T (W_l\oplus W_k) ) $$
given by $\iota_{l,l+k}(\xi)=\xi \otimes 1_{W_k}$ is an embedding of Hilbert spaces.

\begin{definition}
We define $L^2(\cF,\bigwedge^* T\cF )$ as the inductive limit of the Hilbert spaces 
$$(L^2 (\bigoplus_{l=1}^nW_l, \bigwedge^* T\bigoplus_{l=1}^n W_l), \iota_{n,n+1}) ,$$
where the map 
$$ \iota_{n,n+1}:L^2 (\bigoplus_{l=1}^nW_l, \bigwedge^* T\bigoplus_{l=1}^n W_l)\to L^2 (\bigoplus_{l=1}^{n}W_l, \bigwedge^* T\bigoplus_{l=1}^{n+1} W_l)  $$
is defined by 
 $$\iota_{n,n+1}(\xi)=\xi \otimes 1_{W_{n+1}}.$$
\end{definition}

\begin{remark} 
The construction of $L^2(\cF,\bigwedge^* T\cF )$ is, apart from the bundle $\bigwedge^* T\cF$, the same as the construction of  $L^2(\cF)$.
\end{remark}

\section{The holonomy-diffeomorphism algebra and its Hilbert space representations }

In this section we introduce the $\mathbf{HD}(M)$-algebra and its Hilbert space representations.\\

Let $X$ be a vector-field on $M$,  let $\nabla$ be a $\mathfrak{g}$-connection, and let $S$ be a bundle in which $\nabla $ acts.  Denote by $t\to \exp_t(X)$ the corresponding flow. Given $x\in M$ let $\gamma$ be the curve  
$$\gamma (t)=\exp_{t} (X) (x) $$
running from $x$ to $\exp_1 (X)(x)$. We define the operator 
$$e^X_\nabla :L^2 (M , S) \to L^2 (M ,  S )$$
in the following way:
we consider an element $\xi \in L^2 (M ,  S)$ as a function with values in $S$, and define 
\begin{equation}
  (e^X_\nabla \xi )(\exp_1(X) (x))=  ((\Delta \exp_1) (x))  \hbox{Hol}(\gamma, \nabla) \xi (x)   ,
  \label{chopin1}
 \end{equation}
where $\hbox{Hol}(\gamma, \nabla)$ denotes the holonomy of $\nabla$ along $\gamma$ and where $\Delta$ is a factor that secures that $e^X$ is a unitary operator, see \cite{AGnew} for details. 
We then have an operator valued function on $\ca$ defined via 
\begin{equation}
\ca \ni \nabla \to e^X_\nabla  . 
\nn
\end{equation}
We denote this function $e^X$ and call it a holonomy-diffeomorphisms. 
Denote by $\mathscr{F} (\ca , \mathbb{B} (L^2(M, S) ))$ the bounded operator valued functions over $\ca$. This forms a $C^*$-algebra with the norm
$$\| \Psi \| =  \sup_{\nabla \in \ca} \{\|  \Psi (\nabla )\| \}, \quad \Psi \in  \mathscr{F} (\ca , \mathbb{B} (L^2(M, S )) ). $$

\begin{definition}
Let 
$$C =   \hbox{span} \{ e^X | \ X \hbox{ vector field on  } M\}  . $$
The holonomy-diffeomorphism algebra $\mathbf{H D} (M,S,\ca)   $ is defined to be the $C^*$-sub-algebra of  $\mathscr{F} (\ca , \cb (L^2(M,S )) )$ generated by $C$.
We will often denote $\mathbf{H D} (M,S,\ca)   $ by  $\mathbf{H D}  (M)$ when it is clear which $S$ and $\ca$ are meant.
\end{definition}

It was shown in \cite{AGnew} that  $\mathbf{H D} (M,S,\ca)   $ is independent of the metric $g$ on $M$. For further details on the $\mathbf{H D} (M,S,\ca)   $ algebra see \cite{Aastrup:2014ppa}.

\subsection{Representations of the $\mathbf{HD}$-algebra}

In \cite{Aastrup:2017vrm} we proved that a representation of $\mathbf{HD}(M)$ on $$L^2 (\cf ,\bigwedge^* T\cf)\otimes L^2(M,S)  $$ exists provided the real number $p$ in (\ref{regn}) fulfills 
\begin{equation}\frac{1}{d} \left( \frac{d-1}{2}-2p\right)<-\frac12 , \label{betingelse}
\end{equation}
where $d$ is the dimension of $M$.
The way the representation works is the following: If we are given a path $l$ on $M$ we get an operator $h_l : L^2 (\cf)\otimes S_x \to L^2 (\cf)\otimes S_y$ via
$$(h_l\xi) (\nabla)=\hbox{Hol} (l,\nabla )\xi (\nabla) ,$$
where $x$ is the starting point of $l$, $y$ is the endpoint, and $S_x,S_y$ are the fibers of $S$ over $x,y$.
For a general flow $e^X$, where $X$ is an integrable vector field in $M$, we apply the above definition to each of the flows 
$$[0,1]\ni t\to e^{tX}(m) $$
for each $m\in M$.

The problem with this representation is that it requires us to add the factor $L^2(M,S)$ to the Hilbert space. It would be more natural to represent the $\mathbf{HD}$-algebra directly in the fermionic Fock space $\bigwedge^* T\cf$. 
Therefore, in \cite{Aastrup:2018coc} we showed how the $\mathbf{HD}$-algebra can be represented in the Hilbert space $L^2(\cf, \bigwedge^* T\cf)$ without the $L^2(M,S)$ factor. In this representation the same requirement for $p$ holds.

 To explain how the representation in the fermionic Fock space works, we first concentrate on the one particle subspace, i.e. $L^2(\cf, T\cf)$. Let $\nabla$ be a connection in $\cf$ and let $e^X$ be a flow. We thus need an action $e^X_\nabla$ of $e^X$ on $T\cf$. Since $T\cf$ is a subspace of $\Omega^1(M,\mathfrak{g})$ we need to describe how $e^X_\nabla$ move the $\mathfrak{g}$-factor as well as one-forms. For the $\mathfrak{g}$-factor we have the adjoint action of $G$ on $\mathfrak{g}$, therefore the we can take
 $$(e^X_\nabla \xi )(\exp_1(X)(x))=\hbox{Hol}(\gamma , \nabla) \xi (x) \hbox{Hol}(\gamma , \nabla)^{-1}$$
 with $\gamma(t)=\exp (X)(x)$.  We can move the forms with $e^X$ since it is a diffeomorphism, or strictly speaking, since one-forms transforms contravariantly, we move them with $e^{-X}$. 
 
 There are two issues with this construction that are worth noting: First there is no reason why it should preserve $T\cf$. This can however be fixed by taking $T\ca$ instead.
  The second issue is that it is not unitary, not even after multiplying with $(\Delta \exp_1) (x)$ like in (\ref{chopin1}). The reason is that the diffeomorphism $e^X$ in general does not act unitarily on the one-forms. However when the manifold $M$ is compact it is a bounded operator on $T\cf$. If we then extend it multiplicatively to  $\bigwedge^* T\cf $ it will not be a bounded operator anymore, only if we restrict $*$ to a finite set.
  
  The second issue could be remedied by considering $\ca$ to be the space of $SO(3)$-connections on $TM$, or even better spin connections on $M$, since these also act isometrically on $TM$. Hence instead of letting the diffeomorphism $e^X$ act on the one-forms, we let $\mbox{Hol}(\gamma,\nabla)$ act also on the one-forms. This renders the action unitary. Note, however, that this would introduce a metric dependency in the representation. 
 In general we do not see a major problem with a non-unitary representation of $\mathbf{HD}(M)$ that is bounded only on finite-particle states.

  \begin{remark}
  Note that if we chose $\ca$ to be a space of either Levi-Civita or spin connections then we are moving in the direction of a theory of quantum gravity. The interaction between the Dirac operator, which we shall discuss in the next section, and the $\mathbf{HD}$-algebra, encodes the canonical commutation relation between a triad field and a connection related to gravity \cite{Ashtekar:1986yd,Ashtekar:1987gu} (see \cite{Aastrup:2015gba} for details about the canonical commutation relation).
  \end{remark}

\section{The Dirac operator}
\label{Diracoperator}

On each of the subspaces $W_l$ we have a Riemannian metric $\langle \cdot ,\cdot \rangle_{\mathbb{S}^{dim (W_l) +1}}$, where the one-point compactification is isometric to $\mathbb{S}^{dim (W_l) +1}$. We consequently obtain a Dirac operator 
$$D_l:L^2(W_l,\bigwedge^*TW_l) \to L^2(W_l,\bigwedge^*TW_l)  .$$
 
Note that we can  let $D_{l_i}$ act also on   $$L^2 (\bigoplus_{l=1}^nW_l, \bigwedge^* T\bigoplus_{l=1}^n W_l)k$$ when $l_i\leq n$.   
On 
$L^2 (\bigoplus_{l=1}^nW_l, \bigwedge^* T\bigoplus_{l=1}^n W_l)$
we can thus define the operator
$$D_{\oplus n} =\sum_{{l=1}}^n D_l .$$
Since $D_{\oplus n}$ is an elliptic operator on a compact manifold (products of spheres) it is self-adjoint with discrete spectrum.     Note that the function constant equal to $1$ is in the kernel of $D_{\oplus n}$. This allows us to define 
$$D_\cf= \sum_{{l=1}}^\infty D_l ,$$
acting on $L^2(\cF ,\bigwedge^* T\cF)$. Since each  $D_{\oplus n}$ is self-adjoint with discrete spectrum, $D_\cf$ is also self-adjoint and diagonalizable. The spectrum will in general not be discrete due $\cf$ being infinite dimensional.  In particular eigenvalues may have infinite degeneracy.  
Thus $D_\cf$  will not have compact resolvent.

Since we will later need the concrete form of the Dirac operator we will here introduce some notation in order to give this concrete form. 

On $\bigwedge^*T\cF$ we denote by $\mbox{ext} (\varphi)$ exterior multiplication with an element $\varphi \in T\cF$, and by $\mbox{int} (\varphi )$ interior multiplication, i.e.
\begin{eqnarray*}
\mbox{ext} (\varphi )(\psi_1\wedge \ldots \wedge \psi_n)&=& \varphi \wedge \psi_1 \wedge  \ldots \wedge \psi_n \\
\mbox{int} (\varphi )(\psi_1\wedge \ldots \wedge \psi_n)&=&\sum_i(-1)^{i-1} \langle  \varphi, \psi_i\rangle_m \psi_1\wedge \psi_{i-1}\wedge \psi_{i+1}\wedge \ldots \wedge \psi_n .
\end{eqnarray*}
We define Clifford multiplication operators as 
\begin{eqnarray*}
c(\varphi ) &= &\mbox{ext} (\varphi) +\mbox{int} ( \varphi ) \\
\bar{c}(\varphi ) &= &\mbox{ext} (\varphi) -\mbox{int} ( \varphi ) . 
\end{eqnarray*}
If $\{\varphi_i \}$ is an orthonormal basis we have the relations 
$$ \{c(\varphi_i  ),\bar{c}(\varphi_j  )  \}=0, \quad \{c(\varphi_i  ),c(\varphi_j  )  \}=\delta_{ij}, \quad \{\bar{c}(\varphi_i  ),\bar{c}(\varphi_j  )  \}=-\delta_{ij} ,$$
and 
$$c^*(\varphi_i)=c(\varphi), \quad \bar{c}^*(\varphi_i)=-\bar{c}(\varphi) .$$
With this the Dirac operator is locally of the form
$$D_\cf =\sum_{i} \bar{c}(\varphi_i) \nabla_{\varphi_i}^{lc} , $$
where $\{\varphi_i \}$ is a local orthonormal basis in $T\cf$, and where $\nabla^{lc}$ is the Levi-Civita connections.

\subsection{The commutator between the Dirac operator and the $\mathbf{HD}$-algebra }

Let us consider a path $\gamma :[0,1]\to M $, and let $\varphi_i$ be one of the basis vectors in $T_\nabla \cf$, where $\nabla \in \cF$. We can compute
$$\left[ \frac{\partial}{ \partial \varphi_i} , h_\gamma \right] (\nabla ) =\int_0^1 \hbox{Hol} (\gamma_{\leq t}, \nabla) \varphi_i(\dot{\gamma}(t))\hbox{Hol} (\gamma_{\geq t},\nabla )\  dt$$
where $\gamma_{\leq t}$ is $\gamma$ restricted to $[0,t]$ and $\gamma_{\geq t}$ is $\gamma$ restricted to $[t,1]$. Since the holonomies are unital we can estimate
\begin{eqnarray*}
\left\| \left[ \frac{\partial}{ \partial \varphi_i} , h_\gamma \right] (\nabla) \right\| &\leq &
\int_0^1 \left\|  \hbox{Hol} (\gamma_{\leq t}, \nabla) \varphi_i(\dot{\gamma}(t))\hbox{Hol} (\gamma_{\geq t},\nabla) \right\|\  dt \\
&\leq & \int_0^1 \left\|  \hbox{Hol} (\gamma_{\leq t}, \nabla)\| \| \varphi_i(\dot{\gamma}(t)) \| \|\hbox{Hol} (\gamma_{\geq t},\nabla) \right\|\  dt \\
&=& \int_0^1  \| \varphi_i(\dot{\gamma}(t)) \| \   dt =\int_0^1  \| \varphi_i\|_\infty \|\dot{\gamma}(t) \| \   dt  =\| \varphi_i\|_\infty L(\gamma), 
\end{eqnarray*}
where $L(\gamma)$ is the length of $\gamma$, and $\| \varphi_i\|_\infty$ is the uniform norm of $\varphi_i$.

We now focus our attention  on the Dirac operator on just one copy $W_l$. Locally it is of the form 
$$D_l=\sum \bar{c} (e_i)\nabla^{lc}_{e_i} , $$
where $\{e_i\}$ is a local orthonormal frame with respect to the metric on $W_l$, and $\nabla^{lc}$ is the corresponding Levi-Civita connection. Let $k_l$ be the dimension of $W_l$, and let $\{ \varphi_{l_1}, \ldots , \varphi_{l_{k_l}}\}$ be the corresponding basis of $W_l$, and $(x_{l_1}, \ldots , x_{k_{l}})$ the corresponding coordinates. Since we require the $e_i$'s to be normalized we can  asymptotically choose these to be of the form 
$$ e_i= \frac{\varphi_{l_i}}{\| \varphi_{l_1}\|_{\mathbb{S}^{k_l}}}\sim h(x_1,\ldots , x_{k_l}) \varphi_{l_i}    ,$$
with $|h(x_1,\ldots , x_{k_l})|\leq e^{-\frac{1}{2(k_l-2)}}(x_{l_1}^2+\ldots +x_{l_{k_l}}^2)$.
If we therefore want to estimate the commutator between $D_l$ and an $h_\gamma$, we have to estimate 
$$\left\| \left[ \frac{\partial}{ \partial e_i} , h_\gamma \right] (x_{l_1}, \ldots , x_{k_{l}}) \right\|\sim e^{\frac{1}{2(k_l-2)}(x_{l_1}^2+\ldots +x_{l_{k_l}}^2)} \left\| \left[ \frac{\partial}{ \partial \varphi_{l_i}} , h_\gamma \right] (x_{l_1}, \ldots , x_{k_{l}}) \right\| .$$
So this will not be a bounded operator on  $L^2(W_l,\bigwedge^*TW_l)$  due to the factor  $e^{\frac{1}{2k_l}(x_{l_1}^2+\ldots +x_{l_{k_l}}^2)}$. The question is if it is possible to define  the commutator on the vacuum state. To settle this we first estimate 

$$ e^{\frac{1}{2(k_l-2)}(x_{l_1}^2+\ldots +x_{l_{k_l}}^2)} \left\| \left[ \frac{\partial}{ \partial \varphi_{l_i}} , h_\gamma \right]  (x_{l_1}, \ldots , x_{k_{l}}) \right\|  \leq e^{\frac{1}{2(k_l-2)}(x_{l_1}^2+\ldots +x_{l_{k_l}}^2)} \| \varphi_{l_i}\|_\infty L(\gamma )  . $$ 
The commutator with $D_l$ is
\begin{eqnarray*} 
\left[ D_l , h_\gamma \right] ( x_{l_1}, \ldots , x_{k_{l}} ) &\sim & \sum_i \bar{c} (e_i) 
e^{\frac{1}{2(k_l-2)}(x_{l_1}^2+\ldots +x_{l_{k_l}}^2)}  \left[ \frac{\partial}{ \partial \varphi_{l_i}} , h_\gamma \right] (x_{l_1}, \ldots , x_{k_{l}}) . 
\end{eqnarray*}
 Hence we can estimate the expectation value on the vacuum
\begin{eqnarray*}
\langle 0| \left[ D_l , h_\gamma \right] ,\left[ D_l , h_\gamma \right] |0     \rangle  \hspace{-2,8cm}&&
\nn\\
&=& \int_{\R^{k_l}} \langle \left[ D_l , h_\gamma \right] ( x_{l_1}, \ldots , x_{k_{l}} )   , \left[ D_l , h_\gamma \right] ( x_{l_1}, \ldots , x_{k_{l}} ) \rangle  dx_{l_1} \cdots dx_{k_l} \\
&\leq &\frac{1}{\sqrt{\pi}^{k_l}} \int_{\R^{k_l}}  e^{\frac{1}{k_l-2}(x_{l_1}^2+\ldots +x_{l_{k_l}}^2)} \sum_i \| \varphi_{l_i}\|_\infty^2 L(\gamma)^2  e^{-(x_{l_1}^2+\ldots +x_{l_{k_l}}^2)}  dx_{l_1} \cdots dx_{k_l}\\
&=&  \sum_i \| \varphi_{l_i}\|_\infty^2 L(\gamma)^2 \frac{1}{\sqrt{\pi}^{k_l}}  \int_{\R^{k_l}}   e^{-\frac{k_l-3}{k_l-2}(x_{l_1}^2+\ldots +x_{l_{k_l}}^2)}  dx_{l_1} \cdots dx_{k_l}\\  
&=& L(\gamma)^2 \left( \frac{k_l-2}{k_l-3}\right)^{\frac{k_l}{2}} \sum_i \| \varphi_{l_i}\|_\infty^2   
\end{eqnarray*}
This computation can easily be extended to the full Dirac operator. So the condition for the expectation value to exists is $\sum_i \| \varphi_{i}\|_\infty^2<\infty$.

Note that according to \cite{Aastrup:2017vrm} if we choose a $p$ that satisfy the condition (\ref{betingelse}) in section 4.2   $\sum_i \| \varphi_{i}\|_\infty^2$ is finite. Note that we will also be able to define this expectation value as a densely defined operator from  $L^2(\cF,S_{\gamma (0)} \otimes \bigwedge T\cf )$ to $L^2(\cF,S_{\gamma (1)} \otimes \bigwedge T\cf )$, where it is defined on functions having finite support in the first finitely many variables $x_1,x_2,\ldots$ and in the rest of the variables are just the vacuum state. In particular the commutator can be defined on a dense subspace of $L^2(\cF,  \bigwedge T\cf ) \otimes L^2(M,S)$.



\section{An alternative metric}

The advantage of    the metric $\langle \cdot ,\cdot\rangle_m $  constructed in the previous sections  is that it follows directly that the Dirac operator is self-adjoint and in fact diagonalizable.  The downside is that the commutators with the holonomy-diffeomorphism algebra  are unbounded.  In the section we will discuss a different metric, where the commutator between the Dirac operator and the holonomy-diffeomorphism algebra is bounded. 

We set 
$$  f_1(x)=1+e^{-x^2}$$
and 
$$f_k (x)=\frac{\frac{1}{k}+e^{-x^2}}{\frac{1}{k-1}+e^{-x^2}} , \quad k>1 .$$
We define the metric
$$g=\sum_i  \frac{1}{\sqrt{\pi}}\prod_{k=1}^i f_k(x_{k-i}) |\varphi_i \rangle_p  \langle \varphi_i |_p ,$$
where we consider the $(\varphi_i)$'s as vectors in $T_{(x_1,x_2,\ldots )} \cF$, and where the $p$ is the $p$ from the Sobolev norm (\ref{regn}).

Note that 
$$ \prod_{i=1}^\infty  \prod_{k=1}^i f_k(x_{k-i}) =e^{-\sum_{k=1}^\infty x_k^2}, $$
which means that the determinant of this metric is also $\prod_i^\infty \frac{1}{\sqrt{\pi}}e^{-x_i^2}$ as it was the case with the previous metric.

Also, since $f_k (x)\geq \frac{k-1}{k} $ we get
 $$ \prod_{k=1}^i f_k(x_{k-i}) \geq \frac{1}{i} .$$

We thus see that the metric has the following property: In each of the coordinates $x_1,x_2,\ldots$ the metric is not finite, contrary to the metric $\langle \cdot , \cdot \rangle_{m}$. 

Using this metric we can, like in the case of $\langle \cdot , \cdot \rangle_{m}$ construct the Hilbert space $L^2(\cF,\bigwedge^* T\cf )$ as an inductive limit: The $L^2(\cf)$-part is the same construction, since this only depends on the determinant, which in this case is the same.  The construction of the fermionic Fock space is also similar. Since the $|\varphi_i\rangle_p \langle \varphi_i|_p$ part of the metric only depends on coordinates with indices $\leq i$ the map 
$$\iota_{n,n+1}:L^2 (\cf_n , \bigwedge^* T\cf_n) \to L^2 (\cf_{n+1} , \bigwedge^* T\cf_{n+1})  $$
  given by
  $$ \iota_{n,n+1} (\xi)=\xi\otimes 1_{x_{n+1}}$$
  is an embedding of Hilbert spaces.

The Dirac operator we can formally construct as 
$$D_\cf=\sum_{i} \bar{c}(e_i)\nabla_{e_i}^{lc}    ,$$
where $e_i =\frac{\varphi_i}{\|\varphi_i\|_g}$. Note that $\|\varphi_i\|_g$ is a function of  $(x_1,x_2,x_3,\ldots)$ with $\|\varphi_i\|_g\leq \frac{k-1}{k}$. We can thus estimate the commutator with 
\begin{eqnarray*}
\left\|  [D_\cf,h_\gamma ]  \right\| &=& \left\|  \sum_i e_i    \frac{1}{\|\varphi_i \|_g}\left[  \frac{\partial}{\partial \varphi_i},h_\gamma \right]  \right\| \leq  \left\|  \sum_i e_i     i\left[  \frac{\partial}{\partial \varphi_i},h_\gamma \right]  \right\| \\
& \leq & \sqrt{ \sum_i   i^2    \| \varphi_i\|_\infty^2L(\gamma)^2} = L(\gamma)\sqrt{ \sum_i   i^2     \| \varphi_i\|_\infty^2} .
\end{eqnarray*}
Note that $\sum_i   i^2     \| \varphi_i\|_\infty^2$ is in general not finite. However by making the  number $p$ bigger we can arrange for $\sum_i   i^2     \| \varphi_i\|_\infty^2 <\infty$ in which case the commutator will be bounded.

The downside to this metric is that we do not at present know if $D_\cf$ admits a self-adjoint extension. More analysis is required to clarify this issue, in particular we need to analyse the Levi-Civita connection.

\section{The Hilbert space tails}

In this section we will discuss how the construction depends on the parameter $p$ in (\ref{regn}). \\

We will in this section only be concerned with $L^2(\cf )$. In order to see the dependency on $p$ more clearly we make a reformulation: We have defined  $L^2(\cf )$ as the the inductive limit of the Hilbert spaces 
$$ \ch_k=L^2 (\mathbb{R}^k , \pi^{-\frac{k}{2}}e^{-(x_1^2+x_2^2+\ldots +x_k^2)}dx_1\cdots dx_k)$$
with the mappings 
$$ \iota_{k,k+1} (\xi)=\xi \otimes 1,$$
where $1$ is the function constant $1$ in the variable $x_{k+1}$. 

We identify $\ch_k$ with $L^2 (\cf_k)$ since we can identify $\mathbb{R}^k$ with $\cf_k$ via
$$(x_1,\ldots , x_k)\to  x_1\varphi_1+\ldots + x_k\varphi_k .$$

We set 
$$\phi_i =(1+\lambda_i)^p \varphi_i .$$
Note that  $\phi_i$ is an orthonormal basis for $\cf$   with respect to the norm (\ref{regn}) for $p=0$. And we set 
$$\ck_k=L^2 (\R^k, dy_1 \cdots dy_k)  ,$$
and where we think of $\ck_k$ as an $L^2$-space over $\cf_k$, and where $(y_1,\ldots , y_k)$ corresponds to the vector $y_1\phi_1+\ldots +y_k\phi_k$. Note that we can identify $\ch_k$ with $\ck_k $ via the map $\Phi_k:\ch_k \to \ck_k$ given by
\begin{eqnarray*}
\lefteqn{\Phi_k (\xi) (y_1, \ldots , y_k)=}\\
&& \xi \left(  y_1 (1+\lambda_1)^p, \ldots , y_k (1+\lambda_k)^p \right) \left(\prod_{i=1}^k(1+\lambda_i)^p \right)^{-\frac12}
\\&&\hspace{2,2cm}\times 
\pi^{-\frac{k}{4}}e^{-\frac14 (y_1 ^2(1+\lambda_1)^{2p}+ \ldots + y_k^2 (1+\lambda_k)^{2p})} .
\end{eqnarray*}
We can thus identify $L^2(\cF)$ with the inductive limit $(\ck_k,\kappa_{k,k+1} )$ of Hilbert spaces, where  $\kappa_{k,k+1}$ is given by 
$$\kappa_{k,k+1} (\xi)(y_1,\ldots , y_k,y_{k+1})=\xi(y_1,\ldots , y_k) \frac{1}{\sqrt[4]{\pi (1+\lambda_{k+1})^p}}e^{-\frac14 y_{k+1}^2(1+\lambda_{k+1})^{2p}} .$$
The vacuum state in $L^2(\cf)$ is therefore essentially   given by 
$$|0\rangle_p = \prod_{i=1}^\infty\frac{1}{\sqrt[4]{\pi (1+\lambda_{i})^p}}e^{-\frac14 y_{i}^2(1+\lambda_{i})^{2p}} .$$
In the remainder of this section we shall denote $L^2 (\cf)$ by $L^2_p (\cf)$.

If we take $q\not= p$ then due to the "tail"-behaviour of  $|0\rangle_p$ we find first that an easy computation gives   
$$_q\langle 0 |0\rangle_p =0,$$
and second that $L^2_p(\cf )$ is orthogonal to $L^2_q(\cf )$. 
This shows that the construction depends heavily on $p$. Also if we would consider more general forms of the Sobolev-type norm (\ref{regn}), for example a different polynomial of $\Delta$, the construction will be heavily dependent on the polynomial. 

In particular any operator defined on $L^2_p(\cf )$ will not affect $L^2_q(\cf )$. In particular if $p$, or more generally the metric on $\cf$ is dynamic, the dynamic do not stem directly from operators on $L^2_p(\cf )$. In other words: there will be no time-evolution of the tail or the far ultra-violet behaviour of the Hilbert space $L^2_p(\cf )$.

We end the section with the following
\begin{conj}
If $|0\rangle_p$ is orthogonal to $|0\rangle_q$ then the associated representations of the holonomy-diffeomorphism algebra are unitarily inequivalent.
\end{conj}

\section{A BRST operator and a Kasparov type bi-module}
\label{BRST} 

So far we have been concerned with the construction of a spectral triple on the gauge fixing $\cf$. 
In this section we introduce a BRST gauge fixing procedure \cite{Becchi:1975nq,Tyutin:1975qk,Barnich:2000zw,Henneaux:1992ig}, which will permit us to go from the full configuration space $\ca$ to the quotient space $\cf$ in a manner that is compatible with the gauge symmetry.\\

Our first task is to enlarge our construction with a ghost and anti-ghost sector in order to construct a BRST operator. To this end
denote by 
$$
\langle f_1 \vert f_2 \rangle_{\mbox{\tiny $L^2$}} = \int_M \mbox{Tr}(f_1^* f_2)
$$ the inner product between elements $f_i\in L^2(M,\mathfrak{g})$. Denote by $\{f_i\}$ a basis of $ L^2(M,\mathfrak{g})$, which is orthonormal with respect to $\langle \cdot \vert \cdot \rangle_{\mbox{\tiny $L^2$}}$. Consider the fermionic Fock space $\bigwedge^* L^2(M,\mathfrak{g})$. Denote by $\mbox{ext}(f)$ the operator of external multiplication with $f\in L^2(M,\mathfrak{g})$ in $\bigwedge^* L^2(M,\mathfrak{g})$ and 
denote by $\mbox{int}(f)$ its adjoint, i.e. the interior multiplication with $f$:
\begin{eqnarray}
\mbox{ext}(f) (f_1 \wedge  \ldots \wedge f_n) &=&  f\wedge f_1 \wedge  \ldots \wedge f_n,
\nn\\
\mbox{int} (f) (f_1 \wedge  \ldots \wedge f_n) &=& \sum_i (-1)^{i-1} \langle f, f_i \rangle_{\mbox{\tiny $L^2$}} f_1 \wedge \ldots \wedge f_{i-1} \wedge f_{i+1} \ldots \wedge f_n,
\nn
\end{eqnarray}
where $f, f_i\in L^2(M, \mathfrak{g})$. We have the following relations:
\begin{eqnarray}
\{\mbox{ext}(f_1), \mbox{ext}(f_2)  \} &=& 0,
\nn\\
\{\mbox{int}(f_1), \mbox{int}(f_2)  \} &=& 0,
\nn\\
\{\mbox{ext}(f_1), \mbox{int}(f_2)  \} &=& \langle f_1, f_2 \rangle_{\mbox{\tiny $L^2$}}  ,
\end{eqnarray}
as well as
$$
\mbox{ext}(f)^* = \mbox{int}(f),\quad \mbox{int}(f)^* = \mbox{ext}(f),
$$
where $\{\cdot,\cdot\}$ is the anti-commutator. 
Next, we define the ghost and anti-ghost fields
$$
\Theta = \sum_i f_i \otimes\mbox{ext}(f_i)\quad,\qquad \bar{\Theta} = \sum_i f_i \otimes \mbox{int}(f_i).
$$

We use the standard grading, where the ghost has ghost-number equal to one and the anti-ghost has ghost-number equal to minus one.
The aim is to use these additional structures to construct a BRST operator on the complex 
\begin{equation}
\ce=\mathscr{F} (\ca , \cb (L^2(M,S )) ) \otimes \bigwedge^* T\ca  \otimes \bigwedge^* L^2(M,\mathfrak{g}).
\label{complex}
\end{equation} 
%
Next, let $\{{\eta}_i\}$ be yet another complete set of functions in $\OO^1(M,\mathfrak{g})$, which is now orthonormal with respect to the $L^2$-norm. With this we introduce the notation
\begin{eqnarray}
E = \sum_i \frac{\pa}{\pa\eta_i} \otimes   \eta_i  ,
\nn
\end{eqnarray}
where $\frac{\pa}{\pa\eta_i}$ acts on $\ca$ in $\mathscr{F} (\ca , \cb (L^2(M,S )) )$, and $\eta_i\in \OO^1(M,\mathfrak{g}) 
$.
Here 
$E$ represents the field operator\footnote{Note that because $E$ is expressed in terms of a basis that is orthonormal with respect to the $L^2$-norm, it is conjugate to the operator $A$ expressed in the same basis and not to $A$ expressed in the Sobolev basis that we have used until now.}  conjugate to the gauge field $A$.
With this we are now ready to construct the BRST operator
$$
Q = \sum_j W^{(1)}_{j} \otimes \frac{\pa}{\pa {\eta}_j} + W^{(2)} 
$$
where $W^{(1)}_{j}$ and $W^{(2)}$ 
are elements in (\ref{complex}) that all have ghost number equal to one. The $W$'s are determined by the requirement that $Q$ should be nilpotent. We find that this is the case with: 
\begin{eqnarray}
W^{(1)}_{j} &=& \int_M \mbox{Tr} \left(\cd \Theta \star {\eta}_j\right)
\nn\\
W^{(2)} &=& \int_M \mbox{Tr} \left( \bar{\Theta} \Theta^2 \right) dg
\nn
\end{eqnarray}
where $\star$ is the Hodge star, $\cd$ is the covariant derivative acting on the first tensor factor in $\displaystyle{\sum_i f_i\otimes \mbox{ext}(f_i)}$, and where we by $\Theta^2$
mean $\frac{1}{2}\Theta^a \Theta^b \mathrm{i}[T^a, T^b]$ with $T^a$ being the generators of $\mathfrak{g}$ and $\Theta^a= \mbox{Tr}(T^a \Theta)=\sum_i \mbox{ext}(f_i) \mbox{Tr}(T^a f_i)$. Also the integration is over the first tensor factor in  $\displaystyle{\sum_i f_i\otimes \mbox{ext}(f_i)}$. We find:
\begin{equation*}
\left\{ Q, Q \right\} =0.
\end{equation*}

\begin{proof}
First we write
\begin{equation*}
\left\{ Q, Q \right\} =
%
%
%
2\sum_{jk} W^{(1)}_j \left[ \frac{\pa}{\pa {\eta}_j}, W^{(1)}_k  \right]\frac{\pa}{\pa {\eta}_k}
+  2\sum_j \left\{ W^{(1)}_j, W^{(2)}  \right\}\frac{\pa}{\pa {\eta}_j} ,
\end{equation*}
where we used $\{W^{(1)}_j, W^{(1)}_k\}=\{W^{(2)},W^{(2)}\}
=0$. We then compute
\begin{equation*}
\sum_{jk} W^{(1)}_j \left[ \frac{\pa}{\pa {\eta}_j}, W^{(1)}_k  \right]\frac{\pa}{\pa {\eta}_k} 
= \int_M   \mbox{Tr}\left( \Theta^2 \cd\star {E} \right)
\end{equation*}
as well as
\begin{equation*}
\sum_j \left\{ W^{(1)}_j, W^{(2)}  \right\}\frac{\pa}{\pa {\eta}_j} 
=- \int_M \mbox{Tr} (\Theta^2  \cd\star {E} ) 
\end{equation*}
where we used that 
\begin{eqnarray}
\sum_i {\eta}_i(m_1) {\eta}_i(m_2) &=& \d^{(3)}(m_1 - m_2)g \d_{\mathfrak{g}},
\nn\\
\{\Theta(m_1), \bar{\Theta}(m_2)\} &=& \d^{(3)}(m_1-m_2) \d_{\mathfrak{g}},
\nn
\end{eqnarray}
where $\d_{\mathfrak{g}}$ is the Kronecker delta in $\mathfrak{g}$.
This completes the proof.
\end{proof}

\begin{remark}
Note that the proof of $Q$'s nilpotency relies on the integral kernels $\sum_i {\eta}_i {\eta}_i^* $ and $\sum_i f_i f_i^*$ being proportional to the delta function. These integral kernels would, however, have a much more complicated form had we chosen to work with complete sets of functions in $L^2(M,\mathfrak{g})$ and in $\OO^1(M,\mathfrak{g})$ which are orthonormal with respect to the Sobolev-type metric (\ref{regn}) instead of the $L^2$-norm. This means that $Q$ has a highly complex form when expressed in terms of an operator 
${E}$ that is defined in concordance with the Sobolev metric. 
\end{remark}
\begin{remark}
Note that $Q$ depends on a metric $g$ on $M$. This metric dependency comes from the definition of $W^{(2)}$, the Hodge star in $W^{(1)}$, and of course from the choice of the bases $\{f_i\}$ and $\{ \eta_i \}$. 
\end{remark}
We check that the relations
\begin{eqnarray}
&\left[Q  , A  \right]  =  \cd\Theta ,&\quad
\left[Q  , {E}  \right]  = \mathrm{i}[\Theta, {E}]
\nn\\
&\left\{Q  , \Theta  \right\}  = \Theta^2  ,&\quad
\left\{Q  , \bar{\Theta}  \right\}  =\star\left( - \cd\star {E}\right) 
+ 2\mathrm{i}[\bar{\Theta},\Theta]
\nn
\end{eqnarray}
reproduce an odd version of the gauge transformations as is required for the BRST transformation.
We also have the relation 
$$
 \{Q, \mathbf{HD}(M) \}=0,  
$$
which shows that the $\mathbf{HD}(M)$-algebra  lies in the cohomology of $Q$.
Since $Q$ is a nilpotent operator we can as usual consider its zero'th cohomology class $H^0(Q)$. With this we finally arrive at the conjecture:
\begin{conj}
There exist a bijection $\P$ between the  cohomology of $Q$ and the gauge fixing $\cf$, i.e.
$$
\P: H^0(Q)\rightarrow \mathscr{F}(\cf, \cb (L^2(M,S ))) \otimes\bigwedge^* T\ca  .
$$
\end{conj}
We do not provide a proof of this conjecture. It is clear that a proof would require a resolution of the Gribov ambiguity.

\subsection{Kasparov type bi-modules over the $\mathbf{HD}$-algebra}

Let us now focus our attention on the overall mathematical structure that we have obtained. Let us denote by 
$
 \mathscr{F} (\cg)  
$
an algebra of functions on the space of gauge transformations. We shall here not specify precisely what type of functions over $\cg$ we consider.   We have a map  $\ca \to \cg$ by mapping a connection $\nabla$ to the unique element in $g\in \cg$ with $g(\nabla ) \in \cF$ (assuming once more the absence of the Gribov ambiguity).  In this way we can consider $ \mathscr{F} (\cg)$ as an algebra of functions on $\ca$, which are constant along the gauge fixing $\cF$. In particular $ \mathscr{F} (\cg)$ is acting on $\ce$.

We will also extend the action of $D_{\cF}$ to entire $\ca$, and not just $\cf$, since we want $D_\cf$ to act in $\ce$. The way we do this is the following: Since we have assumed that Gribov ambiguity is absent we can write $\ca = \cf \times \cg$, and $D_\cf$  just acts fiberwise in the first coordinate. Note that $D_\cf$ is by construction gauge invariant.

We have
$$
[D_\cf, \mathscr{F} (\cg)]=[\mathbf{HD}(M),\mathscr{F} (\cg)]=0
$$
and thus the complex $\ce$ is a bi-module with a left action of the $\mathbf{HD}$-algebra and a right action of the  $\mathscr{F} (\cg)$-algebra. 
Furthermore, we consider the ghost sector  $\displaystyle{\bigwedge^* L^2(M,\mathfrak{g})}$ in (\ref{complex}) not just as a Hilbert space but also as an algebra, namely the exterior algebra. 
We then have
$$
[D_\cf, \bigwedge^* L^2(M,\mathfrak{g})]_\pm=[\mathbf{HD}(M),\bigwedge^* L^2(M,\mathfrak{g})]=0
$$
where $[\cdot,\cdot]_\pm$ is a graded commutator. 

The inner product in the Hilbert space\footnote{We here include the full fermionic Fock space $\bigwedge^* T\ca$ instead of the gauge fixed Fock space $\bigwedge^* T\cf$, since it is the former that permits a representation of the $\mathbf{HD}(M)$ algebra, see section 4.1. } $L^2(\cf,\bigwedge^* T\ca)$ then turns $\ce$ into a pre-Hilbert $\mathscr{F} (\cg)\otimes \bigwedge^* L^2(M,\mathfrak{g})$-module, since it maps into $\mathscr{F} (\cg)\otimes \bigwedge^* L^2(M,\mathfrak{g})$:
$$
\langle \cdot \vert \cdot \rangle_{L^2(\cf,\bigwedge^* T\ca)} :\ce \times \ce \rightarrow \mathscr{F} (\cg)\otimes \bigwedge^* L^2(M,\mathfrak{g}).
$$
The triple 
$$
(\ce,\mathbf{HD}(M), D_\cf)
$$
then gives us the structure of an unbounded Kasparov bi-module \cite{Kasparov} over $\mathscr{F} (\cg)\otimes \bigwedge^* T_\mathbf{1}\cg$. 

For this to indeed be a Kasparov bi-module  $D_\cf$ would first of all need to have compact resolvent and secondly, its commutator with the $\mathbf{HD}$-algebra would need to be bounded. As we have already seen these two requirements are only partially fulfilled at least for the metric constructed in section 2. In section 9 we will, however, argue that in the case of the alternative metric discussed in section 6, or, more broadly, in the case where we replace the Dirac operator with a Bott-Dirac operator, it is possible that we would have a compact resolvent and therefore that this could be an actual Kasparov bi-module. Compact resolvency is, however, very challenging to prove in the non-Abelian case and thus we are not able to prove this statement.

\begin{remark}
It is tempting to identify the exterior algebra $\bigwedge^* L^2(M,\mathfrak{g})$ with the exterior algebra over the tangent space of $\cg$ in the identity, i.e.
$$
\bigwedge^* L^2(M,\mathfrak{g}) \simeq \bigwedge^* T_\mathbf{1}\cg.
$$
The reason why this identification is not correct is that the gauge transformations in $\cg$ are smooth while the elements in $L^2(M,\mathfrak{g})$ need not be. Morally speaking there is, however, a connection between $L^2(M,\mathfrak{g})$ and $T_\mathbf{1}\cg$ since the ghosts in the former do represent graded infinitesimal gauge transformations. 
This suggest that we should view $\mathscr{F} (\cg)\otimes \bigwedge^* L^2(M,\mathfrak{g})$ as a De Rham complex and the BRST operator $Q$ as an exterior derivative in this complex.
\end{remark}

\section{The Bott-Dirac operator}

In this section we extend the ground state with a complex phase as was first analysed in \cite{Aastrup:2019yui} and in \cite{Aastrup:2020jcf}. There are two reasons for doing this: first, by adding a Chern-Simons term as a complex phase it is possible to build the field strength tensor into a Bott-Dirac type operator whose square then gives us a Yang-Mills Hamiltonian coupled to a fermionic sector. Second, this Bott-Dirac operator has better spectral properties than the Dirac operator introduced in section \ref{Diracoperator}. Note that the addition of a complex phase of this kind is similar to the Kodama ground state known from quantum gravity \cite{Kodama:1988yf,Smolin:2002sz}. Note also that the Bott-Dirac operator, which we construct, is similar to the operator constructed by Higson and Kasparov in \cite{Higson}. \\

We shall here work at the level of the gauge fixing $\cf$.  The first step is to double the construction
\begin{eqnarray}
\mathbf{HD}(M) \longrightarrow \mathbf{HD}^c(M) =  \mathbf{HD}(M)\otimes M_2
\label{HDC}
\end{eqnarray}
together with
\begin{eqnarray}
L^2(\cf,\bigwedge^* T\cf) &\rightarrow& 
\ch^c =
L^2(\cf,\bigwedge^* T\cf) \oplus L^2(\cf,\bigwedge^* T\cf)\nn.
\end{eqnarray}
Next, we write down the Bott-Dirac-type operator
\begin{eqnarray}
B_\cf :=   \left(
\begin{array}{cc}
 D_\cf  &  D_\cf(CS)  \\ 
 -D_\cf(CS)   &  D_\cf
\end{array}
\right),
\label{BDx2}
\end{eqnarray}
where $CS(A)$ is the Chern-Simons functional
\begin{equation}
CS(A) =  \int_M \mbox{Tr} \left( {A}\wedge d{A} + \frac{2}{3} {A}\wedge {A} \wedge {A}\  \right).
\label{CS}
\end{equation}
Note that $B_\cf$ is self-adjoint since $(D_\cf(CS))^*=-D_\cf(CS)$. We find that the state  
\begin{equation}
\ch^c \ni \Phi (A) := \left(
\begin{array}{r}
\cos \left(   CS(A)\right)   
\vspace{0,1cm}
\\
 \sin\left(  CS(A)\right)   
\end{array}
\right) 
\label{platter}
\end{equation}
lies in the kernel of $B_\cf$
$$
B_\cf \Phi =0.
$$
Let us also compute the square of $B_\cf$:
\begin{equation}
B_\cf^2 =
\left(
\begin{array}{cc}
D_\cf^2 - \left(D_\cf(CS)\right)^2  & \{D_\cf, D_\cf(CS)\}
\\
-\{D_\cf, D_\cf(CS)\} & D_\cf^2 - \left(D_\cf(CS)\right)^2
\end{array}
\right).
\label{wein}
\end{equation}
In \cite{Aastrup:2019yui} and \cite{Aastrup:2020jcf} we showed that the diagonal terms in (\ref{wein}) give rise to the Hamilton operator of a Yang-Mills quantum field theory, while the off-diagonal entries give a fermionic term.

\subsection{On compact resolvency}

Let us now take a closer look at the term $D_\cf^2-(D_\cf (CS))^2$. Obviously this depends on what the metric on $\cf$, which is used to construct  $D_\cf$, looks like. We will in the following choose the metric from section 6. This metric is of the form 
$$\sum_i \frac{1}{\sqrt{\pi}} h_i(x_1,\ldots , x_i) |\varphi_i \rangle_p \langle \varphi_i |_p,$$
where $h_i$ are functions with $h_i =\frac{1}{i}$ asymptotically. In particular asymptotically the Dirac operator is of the form 
$$D_\cf  =\sum_i     \bar{c}(e_i) \nabla^{lc}_{e_i}  ,$$
with $e_i=\sqrt{i}\varphi_i $. In particular we expect that in the $D^2_\cf$ term that with increasing $i$ the eigenvalues will grow with $i$. Since $-(D_\cf(CS))^2$ is also a positive term we expect $B_\cf$, modulo the infinity of the fermionic Fock bundle, to have compact resolvent.

We can also consider what happens to the Bott-Dirac operator in the case where we replace the Sobolev norm with the $L^2$-norm and work with the flat metric on $\cf$. The Bott-Dirac operator is constructed like before, but in this case we have
\begin{equation}
D_\cf =\sum_i \bar{c}(\varphi_i ) \frac{\partial }{\pa x_i } ,
\label{stro}
\end{equation}
where now $\{\varphi_i\}_{i}$ is an orthonormal basis for $T_{\nabla_0} \cf $, and we for $\omega \in T_{\nabla_0} \cf $ write 
$$\omega=x_1\varphi_1+x_2\varphi_2+x_3\varphi_3+\ldots  .$$
In this case we have 
$$D_\cf^2 =-\sum_i \frac{\pa^2}{\pa x_i^2}   $$
and 
\begin{equation}
D_\cf (CS)= \sum_i \bar{c}(\varphi_i) \frac{\pa (CS)}{\pa x_i}= \sum_i  2 \bar{c}(\varphi_i)  \int_M \mbox{Tr}\left(\varphi_i\wedge F(A)  \right)
\label{wein2}
\end{equation}
This together with
$$
\sum_i \varphi_i(x)\varphi_i^*(y) = \d^{(3)}(x-y) + \co(\tau)
$$
gives 
\begin{equation}
 - \left(D_\cf(CS)\right)^2  = \int_M \mbox{Tr}\left( F(A)\wedge \star F(A)\right) .
\label{Prigo}
\end{equation}
If we combine (\ref{stro}) with (\ref{Prigo}) we get an expression which in the local limit $\tau\rightarrow 0$ is identical to the Hamilton operator of a Yang-Mills quantum field theory. The off-diagonal terms in (\ref{wein}) can then be interpreted as a fermionic sector. See \cite{Aastrup:2020jcf} and \cite{Aastrup:2017atr} for more details.

In the Abelian case we can rewrite the $F^2$-term in (\ref{Prigo}) as
$$
\int_M \mbox{Tr}\left( F(A)\wedge \star F(A)\right) = \sum_i \lambda_i x_i^2,
$$
where $\{\lambda_i\}$ are the eigenvalues of the Laplace operator on $M$. We thus get 
\begin{eqnarray}
(B_\cf)^2
&=&
\nn\\
&&\hspace{-2cm}
\sum_i  \ \left(
\begin{array}{cc}
 -\frac{\pa^2}{\pa x_i} +   \lambda_i x_i^2   &  -2 \int_M \mbox{Tr} \left(\varphi_i \wedge d \varphi_i  +2\varphi_i \wedge dA \frac{\pa}{\pa x_i}\right)
\\
2 \int_M \mbox{Tr} \left(\varphi_i \wedge d \varphi_i  +2\varphi_i \wedge dA \frac{\pa}{\pa x_i}\right) &  -\frac{\pa^2}{\pa x_i} +   \lambda_i x_i^2  
\end{array}
\right), \nn\\
\label{WE}
\end{eqnarray}
which shows that we have an infinite system of harmonic oscillators on the diagonal, which are each weighted by the eigenvalues of the Laplace operator. This suggests that the Bott-Dirac operator in this case will have a compact resolvent modulo the infinity of the fermionic Fock space at least asymptotically.

\subsection{Alternative formulations}

There are several ways to construct the Bott-Dirac operator, each of which lead to a different expression for the fermionic sector. With the operator (\ref{BDx2}) the fermionic sector gave in the Abelian case an expression that does not actually involve fermionic fields (the off-diagonal entries in (\ref{WE})). To demonstrate how this can be changed we formulate the alternative Bott-Dirac-type operator
\begin{eqnarray}
B'_{\cf} :=   \left(
\begin{array}{cc}
 0  &  D_\cf + \mathrm{i} \overline{D}_\cf(CS)  \\ 
D_\cf - \mathrm{i} \overline{D}_\cf(CS)   &  0
\end{array}
\right),
\label{solpanel}
\end{eqnarray}
where $\overline{D}_\cf$ is defined as
$$
\overline{D}_{\cf}  = \sum_i {c}(\varphi_i) \nabla^{lc}_{\varphi_i}  .
$$
The operator $B'_\cf$ is self-adjoint and has the kernel
$$
\Phi'(A) := \left(
\begin{array}{r}
\exp \left( \mathrm{i}  CS(A)\right)   
\vspace{0,1cm}
\\
 \exp\left(- \mathrm{i} CS(A)\right)   
\end{array}
\right) 
$$
i.e.
$B'_\cf\Phi'(A) =0$. The square of $B'_\cf$ then gives
\begin{eqnarray*}
\left(B'_\cf\right)^2 &=&
\nn\\
&&\hspace{-2cm}
\left(
\begin{array}{cc}
 D_\cf^2 + \left(\overline{D}_{\cf} (CS)\right)^2 - \mathrm{i}[D_\cf,\overline{D}_{\cf} (CS)]  &  0  \\ 
0   &  D_\cf^2 + \left(\overline{D}_{\cf} (CS)\right)^2 + \mathrm{i}[D_\cf,\overline{D}_{\cf} (CS)]
\end{array}
\right).
\end{eqnarray*}
Here we have once more two terms $D_\cf^2 + \left(\overline{D}_{\cf} (CS)\right)^2$ which will give us the Hamiltonian of a Yang-Mills system. In addition to this we have a new fermionic sector given by
\begin{equation}
\mathrm{i}[D_\cf,\overline{D}_{\cf} (CS)] = \sum_{ij} \mathrm{i} c_i \bar{c}_j \left( \left[\nabla^{lc}_{\varphi_i}, \nabla^{lc}_{\varphi_j}(CS)\right] + \left\{\nabla^{lc}_{\varphi_i}, \nabla^{lc}_{\varphi_j}(CS)\right\}  \right).
\label{bobbob}
\end{equation}
If we for simplicity assume that we have the trivial metric on $\cf$, in which case we can write
$
\nabla^{lc}_{\varphi_i} = \frac{\pa}{\pa x_i}
$,
then the first term in the fermionic sector can be written as
$$
\sum_{ij} \mathrm{i} c_i \bar{c}_j \left[\nabla^{lc}_{\varphi_i}, \nabla^{lc}_{\varphi_j}(CS)\right]=2\mathrm{i}\int_M \mbox{Tr}\left(-\bar{\Psi}\nabla^A \Psi + \Psi \nabla^A \bar{\Psi} \right)
$$
where $\nabla^A$ is the covariant derivative and where we used
$$
 \frac{\pa^2 CS}{\pa x_{i } \pa x_{j }} 
 =    \int_M \mbox{Tr} \left(  \varphi_{i  }  \wedge \nabla^A  \varphi_{j} \right) + \int_M \mbox{Tr} \left(  \varphi_{j  }  \wedge \nabla^A  \varphi_{i} \right)   ,   
$$
together with the convention
$$
\Psi = \sum_i \mbox{ext}(\varphi_i)\otimes \varphi_i , \quad \bar{\Psi} = \sum_i \mbox{int}(\varphi_i)\otimes \varphi_i .
$$
The operators $\Psi, \bar{\Psi}$ should be interpreted as quantised fermionic fields since they obey the relation
$$
\{\Psi(m_1), \bar{\Psi}(m_2)\} = \d^{(3)}(m_1-m_2)g \d_{\mathfrak{g}} + \co(\tau).
$$
Thus, the first term in (\ref{bobbob}) looks like a fermionic Hamilton operator.
The second fermionic term 
$ \sum_{ij} \mathrm{i} c_i \bar{c}_j   \left\{\nabla^{lc}_{\varphi_i}, \nabla^{lc}_{\varphi_j}(CS)\right\} $ in (\ref{bobbob}) is, however, less easy to interpret and thus we end the computation here. The point of this exercise is simply to show that one can vary the form of the Bott-Dirac operator in order to obtain various types of fermionic sectors. We refer the reader to \cite{Aastrup:2020jcf} and \cite{Aastrup:2017atr} for more details.

\begin{remark}
In the setup presented here the fermions will a priori have spin-one. The question is whether it is possible to have fermions with half-integer? One option would be to choose the gauge group $G$ to be $SU(2)$ and interpret $S$ as a spin-bundle in which case the fermions would have spin 3/2. Another option was discussed in \cite{Aastrup:2020jcf}, where we formulated a Bott-Dirac operator that gave us spin-half fermions. More work is needed, however, to answer this question definitively.
\end{remark}

\subsection{Modifications and unitary fluctuations}

Let us end this section by noting that it is possible to modify $B_\cf$ in (\ref{BDx2}) with a matrix factor
\begin{eqnarray}
B^U_\cf :=   \left(
\begin{array}{cc}
 D_\cf  &  D_\cf(CS) U   \\ 
 -U^*D_\cf(CS)    &  D^U_\cf
\end{array}
\right)
\label{BDx22}
\end{eqnarray}
where $U$ is a $n$-by-$n$ matrix ($n$ is the dimension of the representation of $G$) that acts in the fermionic Fock space $\bigwedge^* T\cf$ and where $D_\cf^U= U^*D_\cf U$, so that the modified operator has the same kernel as before:
$$
B^U_\cf \Phi =0.
$$
Note that $B^U_\cf$ is still self-adjoint. The square of $B^U_\cf$ then gives
$$
\left(B^U_\cf\right)^2 =
\left(
\begin{array}{cc}
D_\cf^2 - \left(D_\cf(CS)\right)^2 & \{D_\cf, D_\cf(CS)\} U
\\
-U^*\{D_\cf, D_\cf(CS)\} & U^*\left( D_\cf^2  - \left(D_\cf(CS)\right)^2 \right) U
\end{array}
\right).
$$
if we chose $G=SU(2)$ and work with a two-dimensional representation, and if we assume that $U$ is unitary, then we can rewrite it as
$$
U = N \mathds{1}_2 + N^a \sigma^a
$$
where $\sigma^a$ are the Pauli matrices. With this setup it is tempting to interpret $U$ in terms of the lapse of shift fields, that encode the foliation of space and time. This interpretation is motivated by the computations in \cite{Aastrup:2020jcf} that connected the square of a Bott-Dirac operator similar to $B_\cf$ to the Hamilton operators of a Yang-Mills theory coupled to a fermionic sector. 

Moreover, note that $U$ can also include an element in the $\mathbf{HD}$-algebra, in which case $D^U_\cf$ is the Dirac operator that has subjected to inner fluctuations of the algebra. In the case of the standard model such fluctuations are known to give rise to the entire bosonic sector \cite{Connes:1996gi}.

\section{The emergence of a dynamical metric on $M$}

Until now we have been concerned with the construction of metrics and Dirac operators on $\cf$, as well as Hilbert space representations of the $\mathbf{HD}$-algebra. In the following we will discuss to what extend a geometry of the underlying spatial manifold $M$ is encoded in the geometrical construction over $\cf$. The point is that so far we have used a metric on $M$ to construct metrics on $\cf$, but ultimately this ordering should be reversed, so that it is the geometry of $\cf$ that is primary, and the geometry of $M$ that is secondary and emergent. In this section we outline what this will look like.

Thus, let
$g_{\cf}$ be a metric on $\cf$ that satisfies a set of conditions, which we shall not formulate in detail here, but simply state the first of them, which is that it gives rise to a spectral triple $(\mathbf{HD}(M),D_\cf,\ch)$ over $\cf$. This condition guarantees that we have a time-evolution generated by the square of the Dirac operator. 
Now, in the local and flat limit (local and flat with respect to $\cf$) the metric $g_{\cf}(\nabla )$ will give us an $L^2$-metric on $\cf$
$$
 g_{\cf} \stackrel{\mbox{\tiny local limit}}{\longrightarrow}  g_{\cf}^{\mbox{\tiny$L^2$}}
$$
 where $g_{\cf}^{\mbox{\tiny$L^2$}}$ is an $L^2$-norm on $\OO^1(M,\mathfrak{g})$ with respect to a complete set $\{\phi_i\}$ of eigenfunctions of a Laplace operator, which is defined using a metric $g_M$ on $M$. Note that $g_M$ can vary over $\cf$, which means that $g_{\cf}^{\mbox{\tiny$L^2$}}$ also depends on $\nabla$.  

Note that both metrics on $\cf$, which we discussed in section 2 and 6, satisfy this condition. If we insert the parameter $\kappa$ in a manner so that the determinant of the metrics becomes
$$
\exp(-\sum_i x_i^2)\rightarrow \exp(-\frac{1}{\kappa}\sum_i x_i^2)
$$
and if we rename the parameter $\tau$ in the Sobolev type norm (\ref{regn}) to $\kappa$, 
then both metrics will give the $L^2$-metric in the local and flat limit $\kappa\rightarrow 0$. In both of these cases we used, however, the same metric $g_M$ in all points in $\cf$, which means that it will be static, as we shall see in the following. 

But before we discuss time-evolution let us briefly note that it is not difficult to construct metrics on $\cf$ depending on a parameter $\kappa$, which have interesting limits as $\kappa \to 0$. We will for simplicity take a discreet parameter, let us say $\frac{1}{n}$, $n\to \infty$ instead of $\kappa$. If we for example take the metric constructed in section 2 we can now do the following:  

We assume that for each $\nabla  \in \cf$ we have a Riemannian metric $g_\nabla$ on $M$. Furthermore let    $\{\varphi_i \}_{i\in \mathbb{N}}$ be an orthonormal basis with respect to the inner product (\ref{regn}) with $p=0$. We here consider the Sobolev norm with respect to the given metric on $M$, and not the family  $\{ g_\nabla \}$. On the subspace $\hbox{span}\{\varphi_1,\ldots , \varphi_n \}$ we define the family of metrics
$$\langle  \varphi_i, \varphi_j \rangle_{n,\nabla} =\int_M g_\nabla (\varphi_i (m) , \varphi_j(m))dg_\nabla (m)  ,$$
and on $\hbox{span}\{\varphi_{n+1},\varphi_{n+2}\ldots  \}$ we define 
$$ \langle  \varphi_i, \varphi_j \rangle_{n,\nabla} = \langle \varphi_i,  \varphi_j \rangle_m  , $$
where $\langle \cdot , \cdot \rangle_{m}$ is the metric constructed in section 2.2. 
This family of metrics has the following property: 
$$\lim_{n\to \infty} \langle \cdot , \cdot \rangle_{n,\nabla} =g_\nabla .$$
This in particular means that we in the limit can get any family of metrics over $\ca$. There will, however, be restrictions on the family of metrics we can choose, since we would require that we for each $n$ have a spectral triple, and if we also impose that the determinant of the metric should give a vacuum state, i.e. have volume one, there will be conditions on $g_\nabla$ when $\nabla$ is big.

Nevertheless, in the general case the metric $ g_{\cf}(\nabla)  $ will have a dependency on $\nabla$.  In particular it will interact with the Hamiltonian, i.e. we will have a time-evolution of the form 
\begin{equation} g_{\cf}^t = \exp(- i(t_0 - t) B^2_\cf )) 
\co(t_0)
\exp( i(t_0 - t) B^2_\cf )).
\label{time}
\end{equation}
The local limit $g_{\cf}^{\mbox{\tiny$L^2$}}$ will therefore also have a time-evolution. Furthermore, if we consider a state concentrated around a $\nabla$ we will have a time-evolution of the metric   $g_{\cf}^{\mbox{\tiny$L^2$}}(\nabla)$. Thus we arrive at a tentative conclusion that a dynamical metric on $M$ will emerge from a general family of metrics on $\cf$ in a  local limit.

A few remarks: 
\begin{enumerate}
    \item The time-evolution only depends on the metric $g_{\cf}$, in particular it only depends on the choices going into $g_{\cf}$. Thus, if, as it is the case with the constructions in this paper, $g_{\cf}$ does not depend on on chosen coordinate system on $M$, then neither does the time-evolution.
    \item It is not clear if this will generate a four dimensional metric. If this is a theory of emergent gravity it should generate general relativity in the limit $\kappa \to 0$. In this sense, $\kappa$ should probably be related to the Planck's length. 
    \item We are here considering families of metrics, depending on a parameter $\kappa$, and for $\kappa =0$ this should be the classical metric, i.e. general relativity. However as we have seen in this section there does not need to be any connection between the metric in $\kappa =0$ and for a $\kappa \not= 0$. There should however be a connection, since we want the physics to happen for a fixed $\kappa \not=0$, and general relativity should be emergent from the theory. Therefore how $g_{\cf}$ depends on $\kappa$ is of physical importance, and there should be an underlying principle for which dependencies are allowed. For example the metrics constructed in section 2 and 6 will have stationary limits for $\kappa \to 0$, which seems  unphysical in light of general relativity. 
    \item It should of course be checked if the time-evolution of $g_{\cf}$ is well defined, since it is a priory not an observable, i.e. not an operator on the Hilbert space.
\end{enumerate}

\section{Discussion}

In this paper we have shown that a non-perturbative framework, that incorporates the basic building blocks of bosonic and fermionic quantum field theory as well as key elements of general relativity, exists in $3+1$ dimensions. 
Perhaps the most interesting aspect of this framework is the simplicity of its starting point: the spectral triple over the quotient space $\cf$ emerges from an unbounded Kasparov type bi-module over two canonical algebras: the $\mathbf{HD}$-algebra and an exterior algebra of functions on the space of gauge transformations. The $\mathbf{HD}$-algebra simply encode how tensor degrees of freedom are moved around in space and the second algebra encodes the gauge symmetries present in the $\mathbf{HD}$-algebra. What we propose is to analyse the unbounded Kasparov bi-modules over this canonical algebraic setup, and what we find is a framework that possesses the two basic characteristics of a fundamental theory: a high level of canonicity and several layers of unification.\\

This construction involves a set of novel concepts. In the following we will discuss some of the questions that it raises.
Let us begin with the construction of the metrics on the quotient space $\cf$. It is notable that this is done in a two-tier manner: first we construct the Sobolev-type norm on one-forms on $M$, and secondly we use a basis with respect to this norm to construct a metric on $\cf$ together with the corresponding Dirac operator. What this means is that the ultra-violet regularisation, which the Sobolev norm represents, will not be encoded into the spectrum of the Dirac operator on $\cf$. This is in fact a direct consequence of our initial requirement that the volume of the quotient space $\cf$ should be finite. If we had constructed the metric on $\cf$ in a way so that each dimension (which correspond to a degree of freedom of a gauge field) got successively 'smaller' then the volume of $\cf$ would be zero and hence the Hilbert space inner product would vanish.

Now, this seems to be somewhat in contradiction to our initial claim that both the metric on $\cf$ and the ultra-violet regularisation, which it gives rise to, are dynamical: if the regularisation represented by the Sobolev norm is not encoded into the spectrum of the Dirac operator, then how can it be dynamical? This is a technical question, which must be answered in several steps. First of all, the time-evolution generated by the Dirac operator will always be confined to take place within a given Hilbert space. The choice of the Sobolev norm is essentially a choice of such a Hilbert space and a representation of the $\mathbf{HD}$-algebra. That means that the Hilbert space itself and its far-ultra-violet tails cannot be shifted by means of the time-evolution. What can be shifted are {\it finite} parts of states corresponding to finitely many variables. In other words, it is not {\it all} of the regularisation that can be subjected to time-evolution, only finite (but arbitrarily large) parts of it.  

It is possible that the finite-volume assumption should ultimately be eased. It seems intuitively appealing that the ultra-violet regularisation should be encoded in the spectrum of the Dirac operator. Note, however, that if we do that then the determinant of the metric will no longer regularise the Hilbert space in a straight-forward manner.

In the case of the metric in section 2 the Dirac operator is self adjoint, but the commutator with the $\mathbf{HD}$-algebra is unbounded, and hence this does not directly give rise to a spectral triple. If we however choose the alternative metric from section 6 the commutator is bounded. In that case, however, further analysis of the Dirac operator is necessary in order to determine whether it is self adjoint. In particular we need an analysis of the associated Levi-Civita connection. This is also needed to development a smooth calculus for the Dirac operator.

Concerning the Hilbert space tail, which is the ultra-violet limit of the ground state in the Hilbert space, then it is an interesting question what physical implications it might have. In terms of scale these tails exist at infinitely high energy scales and as such it seems unlikely that they can have any great physical implications. On the other hand, each tail represents a unitarily inequivalent representation and thus one would expect there to be some tangible difference between the different representations. Another interesting question is whether there could exist singular points in the time-evolution, where phase transitions between different unitarily inequivalent representations are possible. Specifically, one might think that the big bang could be such a point. This would make sense since the limit where it becomes possible to shift between different representations is precisely the infinite-energy limit. Note that these questions concerning the Hilbert space tails are closely related to the question to what extend our construction is background independent.

With respect to the Sobolev-type norm then it is interesting that it only plays a role in conjunction with the $\mathbf{HD}$-algebra. If we only had the Dirac operator we could ignore the Sobolev norm altogether. It is when we seek a representation of the $\mathbf{HD}$-algebra, which requires a identification of gauge field operators, that the Sobolev norm plays a central role. This feature is also reflected in the observation that the Bott-Dirac operator appears to have compact resolvent also in the case when all the regularisation, which the metric on the quotient space provides, is turned off. This interplay between the Dirac and Bott-Dirac operators, the $\mathbf{HD}$-algebra, and the question of regularisation is intriguing and should be analysed in greater detail.

Concerning the dynamical ultra-violet regularisation, then this concept appears to be much more general than our current application: whenever there is an ultra-violet regularisation in a non-perturbative quantum gauge theory it must be dynamical if it is to be compatible with the gauge symmetry, and once it is dynamical it can be interpreted as a physical feature and not merely a computational artefact. What does this actually mean? It seems to suggest that there could be an anti-gravitational forcing that would not only put into question the necessity of quantising gravity but which could also have cosmological implications. If a regularisation can be interpreted as being physical (say, at the Planck scale), then it must play a role in for instance the black hole and big bang singularities. 

The emergence of a dynamical metric on the underlying manifold $M$ is probably the most significant result i this paper. This result raises a few questions. First of all, will a four-dimensional metric emerge from this and if yes, what will the signature of that metric be? At the moment we have not been able to answer this question although it should be possible to do so with the information available. Another question comes from the dependency of the metric on $A$, which gives rise to the time-evolution. At the moment it is possible to have a static metric on $M$ (i.e. no $A$-dependency) with an arbitrary geometry. For instance, this is the case with the metrics we constructed in section 2 and 6. This is clearly not physically acceptable and thus one might wonder whether there exist a mechanism so that only the flat metric on $M$ can be static. One possible answer could be that the configuration space $\ca$ is a space of $SO(3)$ Levi-Civita connections.


One of the challenges in this approach is to understand how spin-half fermions may emerge. In a previous publications we devised a method to incorporate spin-half fermions but this came at the price of injecting some {\it ad hoc}-ness into the construction. An alternative solution could be to choose the gauge group to be $SU(2)$ and then interpret the bundle, in which the group acts, as the spin-bundle. This would first of all mean that the fermions have spin $3/2$, secondly it would increase the level of canonicity to our construction, since the choice of gauge group is a key initial input, and thirdly it would cast our construction in the light of a quantum gravitational theory.

With respect to interpretation of our construction it seems clear that the framework should be interpreted as a candidate for a Planck-scale theory. The question is what theory it will produce in the local limit, which is the limit where (ill defined) perturbative quantum field theory emerges, and in the semi-classical limit. Concerning the latter it is worth noting that the $\mathbf{HD}$-algebra produces a matrix algebra in the limit where states in the Hilbert space a localised around a single classical point. The question, which started off this research project, is whether this matrix algebra could be related to the noncommutative formulation of the standard model \cite{Connes:2006qj,Chamseddine:1991qh,Chamseddine:2006ep,Farnsworth:2014vva,Boyle:2016cjt} (see \cite{Chamseddine:2019fjq} for an interesting historical overview) due to Chamseddine and Connes?

Finally it is important to note that throughout our paper we have assumed the absence of the Gribov ambiguity \cite{Gribov:1977wm}. This means that we have assumed that it is possible to chose a gauge fixing condition that intersects each gauge orbit only once. However, as Singer pointed out \cite{Singer:1978dk}, this is never possible in a non-Abelian gauge theory.  
The Gribov ambiguity is determined by the kernel of the Faddeev-Popov operator, which must have an empty kernel in order for the Gribov ambiguity to be absent. What Singer found is that it is impossible to choose a gauge fixing condition for a non-Abelian gauge theory so that this condition is met. 
In general it is, however, known that the kernel of differential operators are not very stable objects. This is why one studies the kernel minus the co-kernel in index theory instead of just the kernel, as the former combination is much more stable. This raises, therefore, the question how stable the kernel of the Fadeev-Popov operator is as we vary $A$ throughout the configuration space $\ca$? And in particular it raises the question what measure that kernel has with respect to the Hilbert space that we have constructed? If the kernel of the Faddeev-Popov operator has measure zero then the Gribov ambiguity will be immaterial.

\vspace{1cm}
\noindent{\bf\large Acknowledgements}\\

\noindent
JMG would like to express his sincere gratitude to the entrepreneur Kasper Gevaldig, Copenhagen, Denmark, for his generous financial support. JMG would also like to thank the following sponsors:  Frank Jumppanen Andersen, Bart De Boeck, Simon Chislett, Trevor Elkington, Jos Gubbels, Claus Hansen, Jens Haukohl, Kasper K\o ppen, Hans-J\o rgen Mogensen, Stephan M\"{u}hlstrasser, Ben Tesch and Vladimir Zakharov, as well as the company Providential Stuff LLC. JMG would also like to express his gratitude to the Institute of Analysis at the Gottfried Wilhelm Leibniz University in Hannover, Germany, for kind hospitality during numerous visits.


\begin{thebibliography}{99}




\bibitem{Aastrup:2020jcf}
J.~Aastrup and J.~M.~Grimstrup,
``The metric nature of matter,''
J. Geom. Phys. \textbf{171} (2022), 104408.


\bibitem{Aastrup:2005yk}
J.~Aastrup and J.~M.~Grimstrup,
``Spectral triples of holonomy loops,''
Commun. Math. Phys. \textbf{264} (2006), 657-681.




\bibitem{Connes:2008vs}
A.~Connes,
``On the spectral characterization of manifolds,''
J. Noncommut. Geom. \textbf{7} (2013) no.1, 1-82.




\bibitem{Connes:1996gi}
A.~Connes,
``Gravity coupled with matter and the foundation of non-commutative
geometry,''
Commun.\ Math.\ Phys.\  {\bf 182} (1996) 155.





\bibitem{ConnesBook}
A.~Connes,
``Noncommutative Geometry,'' Academic Press, 1994.







\bibitem{AGnew}
  J.~Aastrup and J.~M.~Grimstrup,
  ``C*-algebras of Holonomy-Diffeomorphisms and Quantum Gravity II'',
   J.\ Geom.\ Phys.\  {\bf 99} (2016) 10.






\bibitem{Aastrup:2017atr}
  J.~Aastrup and J.~M.~Grimstrup,
  ``Nonperturbative quantum field theory and noncommutative geometry,''
  J.\ Geom.\ Phys.\  {\bf 145} (2019) 103466.



\bibitem{Aastrup:2019yui}
J.~Aastrup and J.~M.~Grimstrup,
``Non-perturbative Quantum Field Theory and the Geometry of Functional Spaces,''
Fortsch. Phys. \textbf{69} (2021) no.10, 2100106.









\bibitem{Connes:2006qj}
A.~Connes and A.~H.~Chamseddine,
``Inner fluctuations of the spectral action,''
J. Geom. Phys. \textbf{57} (2006), 1-21.




\bibitem{Chamseddine:1991qh}
A.~H.~Chamseddine and A.~Connes,
``Universal formula for noncommutative geometry actions: Unification of
gravity and the standard model,''
Phys.\ Rev.\ Lett.\  {\bf 77} (1996) 4868.








\bibitem{Chamseddine:2006ep}
  A.~H.~Chamseddine, A.~Connes and M.~Marcolli,
  ``Gravity and the standard model with neutrino mixing,''
  [arXiv:0610241].





\bibitem{Becchi:1975nq}
C.~Becchi, A.~Rouet and R.~Stora,
``Renormalization of Gauge Theories,''
Annals Phys. \textbf{98} (1976), 287-321.




\bibitem{Tyutin:1975qk}
I.~V.~Tyutin,
``Gauge Invariance in Field Theory and Statistical Physics in Operator Formalism,''
[arXiv:0812.0580 ].

\bibitem{Barnich:2000zw}
G.~Barnich, F.~Brandt and M.~Henneaux,
``Local BRST cohomology in gauge theories,''
Phys. Rept. \textbf{338} (2000), 439-569.

\bibitem{Henneaux:1992ig}
M.~Henneaux and C.~Teitelboim,
``Quantization of gauge systems,''
Princeton University Press.






\bibitem{Gribov:1977wm}
V.~N.~Gribov,
``Quantization of Nonabelian Gauge Theories,''
Nucl. Phys. B \textbf{139} (1978), 1.



\bibitem{Singer:1978dk}
I.~M.~Singer,
``Some Remarks on the Gribov Ambiguity,''
Commun. Math. Phys. \textbf{60} (1978), 7-12.








\bibitem{Feynman:1981ss}
  R.~P.~Feynman,
  ``The Qualitative Behavior of Yang-Mills Theory in (2+1)-Dimensions,''
  Nucl.\ Phys.\ B {\bf 188} (1981) 479.

 
\bibitem{Singer:1981xw}
  I.~M.~Singer,
  ``The Geometry of the Orbit Space for Nonabelian Gauge Theories. (Talk),''
  Phys.\ Scripta {\bf 24} (1981) 817.


\bibitem{Orland:1996hm}
  P.~Orland,
  ``The Metric on the space of Yang-Mills configurations,''
  [arXiv:9607134].




\bibitem{Kasparov}
G.~G.~Kasparov, 
``The operator K-functor and extensions of C*-algebras'', 
Izv. Akad. Nauk SSSR Ser. Mat., {\bf 44}:3 (1980), 571–636; Math. USSR-Izv., {\bf 16}:3 (1981), 513–572.



\bibitem{Kodama:1988yf}
  H.~Kodama,
  ``Specialization of Ashtekar's Formalism to Bianchi Cosmology,''
  Prog.\ Theor.\ Phys.\  {\bf 80} (1988) 1024.



\bibitem{Smolin:2002sz}
  L.~Smolin,
  ``Quantum gravity with a positive cosmological constant,''
  [arXiv:0209079].


  
\bibitem{Higson}
N.~Higson and G.~Kasparov, "E-theory and KK-theory for groups which act properly and isometrically on Hilbert space", 
Inventiones Mathematicae, vol. {\bf 144}, issue 1, pp. 23-74.





\bibitem{Aastrup:2014ppa}
J.~Aastrup and J.~M.~Grimstrup,
``The quantum holonomy-diffeomorphism algebra and quantum gravity,''
Int. J. Mod. Phys. A \textbf{31} (2016) no.10, 1650048.








\bibitem{Aastrup:2017vrm}
  J.~Aastrup and J.~M.~Grimstrup,
  ``Representations of the Quantum Holonomy-Diffeomorphism Algebra,''
  [arXiv:1709.02943].

\bibitem{Aastrup:2018coc}
J.~Aastrup and J.~M\o{}ller Grimstrup,
``On the Fermionic Sector of Quantum Holonomy Theory,''
[arXiv:1810.00157].




\bibitem{Ashtekar:1986yd}
  A.~Ashtekar,
  ``New Variables for Classical and Quantum Gravity,''
  Phys.\ Rev.\ Lett.\  {\bf 57} (1986) 2244.

\bibitem{Ashtekar:1987gu}
  A.~Ashtekar,
  ``New Hamiltonian Formulation of general relativity,''
  Phys.\ Rev.\  D {\bf 36} (1987) 1587.




\bibitem{Aastrup:2015gba}
J.~Aastrup and J.~M.~Grimstrup,
``Quantum Holonomy Theory,''
Fortsch. Phys. \textbf{64} (2016) no.10, 783-818.








\bibitem{Farnsworth:2014vva}
S.~Farnsworth and L.~Boyle,
``Rethinking Connes' approach to the standard model of particle physics via non-commutative geometry,''
New J. Phys. \textbf{17} (2015) no.2, 023021.


\bibitem{Boyle:2016cjt}
L.~Boyle and S.~Farnsworth,
``A new algebraic structure in the standard model of particle physics,''
JHEP \textbf{06} (2018), 071.


\bibitem{Chamseddine:2019fjq}
A.~H.~Chamseddine and W.~D.~Van Suijlekom,
``A survey of spectral models of gravity coupled to matter,''
[arXiv:1904.12392 ].
















































  








\end{thebibliography}
\end{document}